\documentclass[twocolumn,showpacs,preprintnumbers,amsmath,amssymb,superscriptaddress]{revtex4-2} 
\usepackage[english]{babel}
\usepackage{multirow}
\usepackage{placeins}
\usepackage{color}
\usepackage{amsmath,amsthm}
\usepackage{graphicx}
\usepackage{comment}
\usepackage{tikz}
\usepackage{bbold} %
\usetikzlibrary{decorations.pathreplacing}
\usepackage{siunitx}
\usepackage{hyperref}

\usepackage[normalem]{ulem}

\hypersetup{
colorlinks = true,
urlcolor   = blue,
linkcolor  = blue,
citecolor  = blue
}

\begin{document}
\title{Driving collective current excitations using light: The two-time $GW$ approach}

\author{Chin Shen Ong}
\affiliation{Department of Physics and Astronomy, Uppsala University, Box-516, SE 75120, Sweden}

\author{Denis Gole\v z}
\affiliation{Jozef Stefan Institute, Jamova 39, SI-1000 Ljubljana, Slovenia}
\affiliation{Faculty of Mathematics and Physics, University of Ljubljana, Jadranska 19, 1000 Ljubljana, Slovenia}

\author{Angel Rubio}
\affiliation{Max Planck Institute for the Structure and Dynamics of Matter, Luruper Chaussee 149, 22761 Hamburg, Germany}
\affiliation{Center for Computational Quantum Physics, Flatiron Institute, Simons Foundation, New York City, NY 10010, USA}
\affiliation{Nano-Bio Spectroscopy Group, Departamento de F\'isica de Materiales, Universidad del Pa\'is Vasco, 20018 San Sebastian, Spain}

\author{Olle Eriksson}
\affiliation{Department of Physics and Astronomy, Uppsala University, Box-516, SE 75120, Sweden}
\affiliation{Wallenberg Initiative Materials Science for Sustainability (WISE), Uppsala University, 75121 Uppsala, Sweden }

\author{Hugo U. R. Strand}
\affiliation{School of Science and Technology, Örebro University, SE-701 82 Örebro, Sweden}

\date{\today}

\begin{abstract}
We identify a distinct transverse collective excitation, which we name the \emph{curron}, arising from current-current interactions in a driven quantum metal. Unlike plasmons, which involve longitudinal charge oscillations, currons are transverse current-density oscillations resulting from the interplay between the vector potential generated by the current and the external driving field. We demonstrate the emergence of this excitation in sodium metal by solving the Kadanoff-Baym equations on a complex time contour within the non-equilibrium two-time (TT) $GW$ formalism, marking, to our knowledge, the first TT-$GW$ calculation on a realistic material. We further show that two-time quantum memory effects leave measurable signatures: a pump-induced elevation in the baseline of the current-to-field response, potentially observable in polarization- and momentum-resolved conductivity experiments. By extracting effective resistive and memory coefficients from the TT-$GW$ dynamics, we introduce a generalized d'Alembert wave equation that captures the many-body damping and retardation inherent to driven quantum systems. These results establish current-current response functions as a platform to harness qualitatively new collective dynamics in correlated matter, opening new avenues for probing light-matter interactions beyond charge-density dynamics.
\end{abstract}

\maketitle

\newpage
\section{Introduction}

In condensed matter physics and nanophotonics, plasmons have emerged as a central paradigm for understanding and manipulating light-matter interactions at the atomistic and nanoscale. Plasmons are collective oscillations of electrons in a material that drive numerous technologies and practical applications, including surface-enhanced Raman scattering (SERS)~\cite{Moskovits1985}, plasmonic waveguides for subwavelength light localization~\cite{Schuller2010}, and advanced biosensing techniques that exploit electromagnetic field enhancements near metallic surfaces~\cite{Altug2022}. On the theoretical front, plasmons have provided deep insights into electron correlation~\cite{Pines1956}, nonlocal response~\cite{Pitarke2007}, and the interplay between quantum and electromagnetic degrees of freedom~\cite{Tame2013}. Their importance thus spans both practical applications (e.g., next-generation optoelectronic and photonic devices~\cite{Ekmel2006}) and fundamental science (e.g., many-body theory of metals and semiconductors~\cite{Hedin1970}).

\begin{figure}[t]
    \centering
    \includegraphics[width=\linewidth]{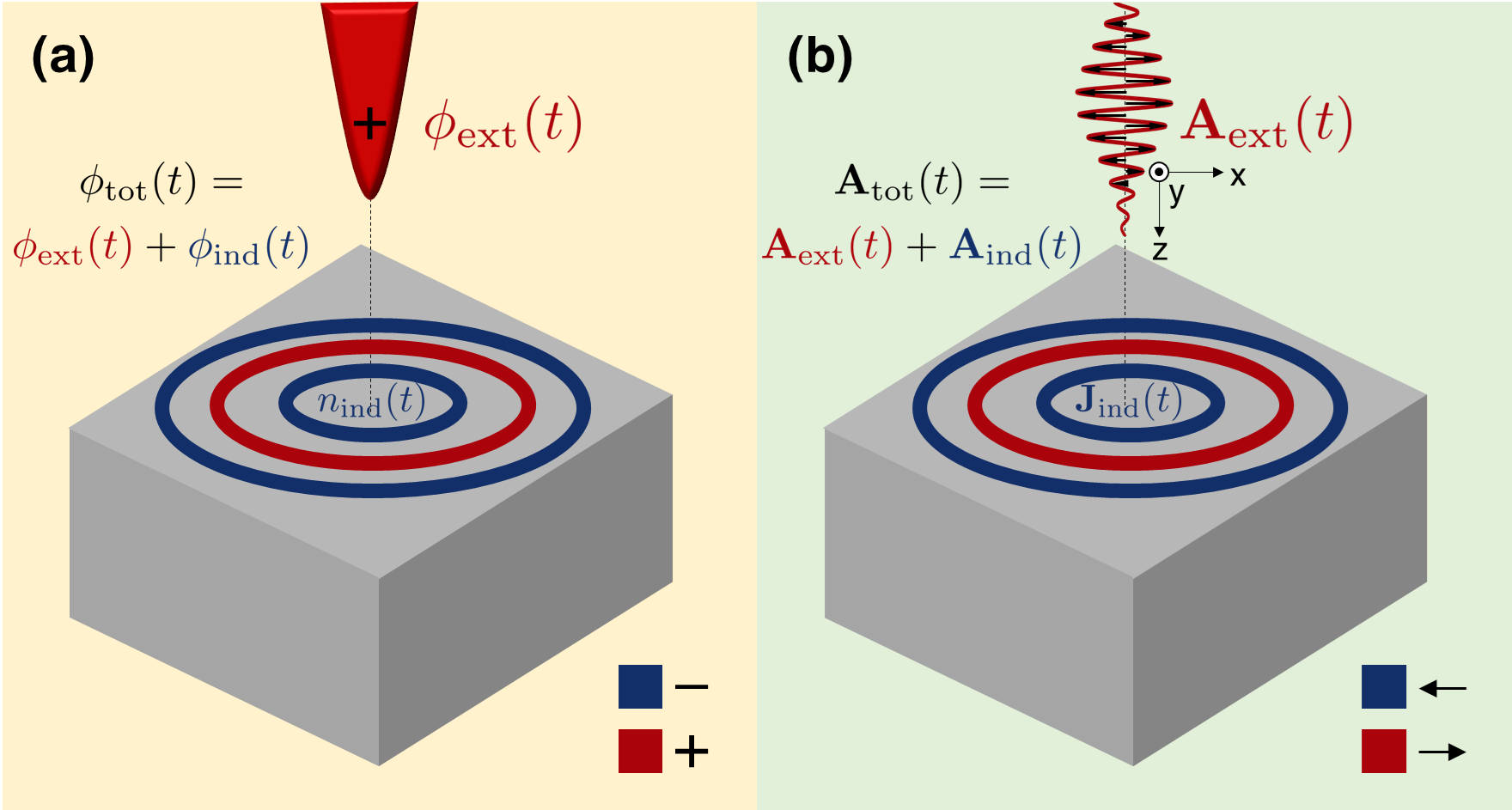}
    \caption{Schematic picture of: (a) electric potential, $\phi(t)$, and a surface plasmon, (b) vector potential, $\mathbf{A}(t)$, and a surface curron.}
    \label{fig:plasmon-vs-curron}
\end{figure}

Physically, the plasmon is excited when the induced scalar potential of the electron fluid (or plasma) feeds back strongly into the total scalar potential at the natural frequency of the electron density, see Fig.~\ref{fig:plasmon-vs-curron}a. This creates a resonance at which the electrons oscillate collectively as a quasiparticle known as the plasmon. Mathematically, the plasmon is excited at the frequency at which the longitudinal dielectric function has a pole, and diverges. This pole signifies that the electronic system can sustain collective density oscillations with long lifetimes, even for small external drives.

Bulk plasmons are longitudinal charge-density oscillations and cannot be excited directly by transverse electromagnetic waves such as light. Nevertheless, early optical studies of alkali metals reported spectral features close to the bulk plasmon frequency.  In the late 1960s, Sutherland \emph{et al.}~\cite{Sutherland1967,Sutherland1969} measured the optical reflectance of sodium films grown on quartz under ultra-high vacuum and extracted the transverse dielectric function, $\varepsilon_{T}(\omega)$.  They interpreted the resonance in $\varepsilon_{T}(\omega)$ as indirect evidence of the bulk plasmon, because in the long-wavelength limit the pole of the transverse dielectric function coincides with the plasmon frequency~\cite{Nozieres1959}. Still, this interpretation raises a deeper question: what physical excitation actually produced the observed resonance?

The clearest distinction between the longitudinal and transverse modes was made by Bohm and Pines themselves, who in the early 1950s introduced the concept of ``collective" excitations in their series of seminal work~\cite{1Bohm1951, 2Pines1951, 3Bohm1953, 4Pines1953}. There, the concept of ``plasma oscillation" was clearly separated into ``longitudinal oscillation" and ``transverse oscillation". In Ref.~\cite{1Bohm1951}, they made it clear that the former originates from the long-range correlation of electron positions brought about by Coulomb interactions, whereas the latter are due to electromagnetic waves being strongly modified by the fields arising from the collective particle response. In a follow-up review by Pines~\cite{Pines1956}, wherein the former was given the name of a ``plasmon", the latter was never named, even though its distinction from the former was explicitly emphasized. Later works calculated the $\varepsilon_{T}(\omega)$ for the free-electron gas~\cite{stern1963} and a model semiconductor~\cite{Sharma1981}, confirming that at zero wave vector, $\varepsilon_{T}(\omega)$ matches the longitudinal dielectric function, $\varepsilon_{L}(\omega)$, even though $\varepsilon_{T}(\omega)$ exhibits distinct behavior at finite wave vectors relevant for optical response, as predicted by Nozi{\`{e}}res and Pines~\cite{Nozieres1959}. In 1969, the feedback onto the electromagnetic field created by the electronic currents was explicitly incorporated for the first time when discussing the optical response of discrete multi-level atomic systems~\cite{Crisp1969}. 

Despite these clarifications, the resulting transverse collective mode is frequently labeled a “transverse plasmon” in the literature. This misnomer is a historical carry-over from earlier spectroscopy and optical studies, where collective electron oscillations, regardless of their transverse or longitudinal character, were loosely termed plasmons. Ritchie’s pioneering 1957 work~\cite{Ritchie1957} on surface plasmons (Fig.~\ref{fig:plasmon-vs-curron}a) was perhaps the earliest implicit appearance of what would later be called a “transverse plasmon.” 
Surface plasmons involve surface charge-density oscillations that are coupled to electromagnetic fields with transverse-magnetic (TM) polarization. The TM label refers to the character of the accompanying field, not to the charge oscillation itself. Ritchie’s work was subsequently described as demonstrating “transverse plasmons” on a metallic surface (e.g., \cite{Castanie2013}), even though he never used that term. Although a surface plasmon-polaritron is accompanied by a TM-polarized electromagnetic field bound to the interface, the oscillating surface charge density is driven by the electric field component normal to the interface, which couples to charge via the discontinuity condition of Gauss’s law.

In this work, we formalize the concept of bulk ``transverse oscillation" as a quasiparticle, naming it the ``curron'', as the analog of the plasmon for collective \textit{current} excitations driven by \textit{vector potentials} rather than charge excitation driven by scalar potentials. Just as plasmons manifest when the induced charge densities reinforce the total scalar potential, currons manifest when the \textit{induced currents} reinforce the total \textit{vector potential} to create a resonant mode~\cite{1Bohm1951,Crisp1969}. Since electromagnetic waves inherently involve transverse fields encoded in the vector potential, by examining how these fields couple to the electron fluid, we aim to uncover the possibility of collective current excitations, or ``currons", at resonance conditions corresponding to poles of the transverse dielectric function~\cite{Sharma1981}, thereby broadening the conceptual framework of collective excitations in electronic systems.

We investigate the excitation of currons in alkali metals such as Na and K. These metals feature nearly spherical Fermi surfaces that closely resemble those of a homogeneous (i.e., spatially uniform) electron gas, making them ideal prototypes for exploring collective charge excitations. Bulk plasmons are well-characterized collective oscillations in these materials, making them a natural candidate for a curron analysis. In the latter part of our work, we incorporate el-el (el-el) interactions into the study via the time-dependent two-time (TT) $GW$ method. In this framework, the density functional theory (DFT) electronic structure is often used as an approximation of the noninteracting system at equilibrium. Since the local-density approximation (LDA) within DFT models the exchange-correlation potential using the homogeneous electron gas as a reference, the Kohn-Sham eigenvalues and wavefunctions of Na is an excellent starting point for subsequent $GW$ corrections. Indeed, the $GW$ quasiparticle bandstructure has been shown to accurately reproduce experimental angle-resolved photoemission spectra~\cite{Hedin1970, Hybertsen1986, Jensen1985, Northrup1987, Lyo1988, Northrup1989}, making Na a prime testbed for assessing the accuracy of the TT-$GW$ approach.

A key feature of the TT-$GW$ implementation presented here, is that we solve for the full nonequilibrium, two-time $GW$ self-energy along the Kadanoff-Baym contour. This approach retains the complete temporal structure of the interacting system, and avoids reducing the two-time dependence to a single time argument through approximations such as the generalized Kadanoff-Baym ansatz (GKBA)~\cite{Lipavsky1986, Kalvova2019, Schlunzen2020} or the adiabatic $GW$ approximation~\cite{Attaccalite2011, Chan2021}. To the best of our knowledge, this work represents the first application of a full nonequilibrium TT-$GW$ formalism to model a realistic system with parameters derived entirely from first-principles calculations.

In this investigation, we confine our analysis to bulk collective modes and defer treatment of surface and interface-related effects involving dielectric discontinuities to future work. We further restrict the analysis to homogeneous $s$-polarized light as the excitation source. The external electromagnetic driver is represented by an applied vector potential, $\mathbf{A}_{\mathrm{ext}}(t)$, which corresponds to an electric field $\mathbf{E}_{\mathrm{ext}}(t)$. For clarity, we define the propagation direction of light to be in the $z$-direction, and polarization of $\mathbf{A}_{\mathrm{ext}}(t)$ to be along the $x$-axis (as shown in Fig.~\ref{fig:plasmon-vs-curron}b), such that
\[
\mathbf{A}_{\mathrm{ext}}(t) = [A_{\mathrm{ext}}^x(t),\, 0,\, 0].
\]
Henceforth, unless otherwise specified, the $x$-components of $\mathbf{A}$, $\mathbf{E}$, and $\mathbf{J}$ are denoted simply as $A$, $E$, and $J$, respectively. The subscripts of ``ext", ``ind" and ``tot" refer to external, induced and total fields/currents, respectively. 
Throughout this paper, we adopt the time-harmonic convention \( e^{-i\omega t} \) (with a negative sign in the exponent) for all time-dependent quantities, and all physical quantities are understood to correspond to the real part of the complex expression unless otherwise stated.

\begin{figure*}
    \centering
    \includegraphics[width=\linewidth]{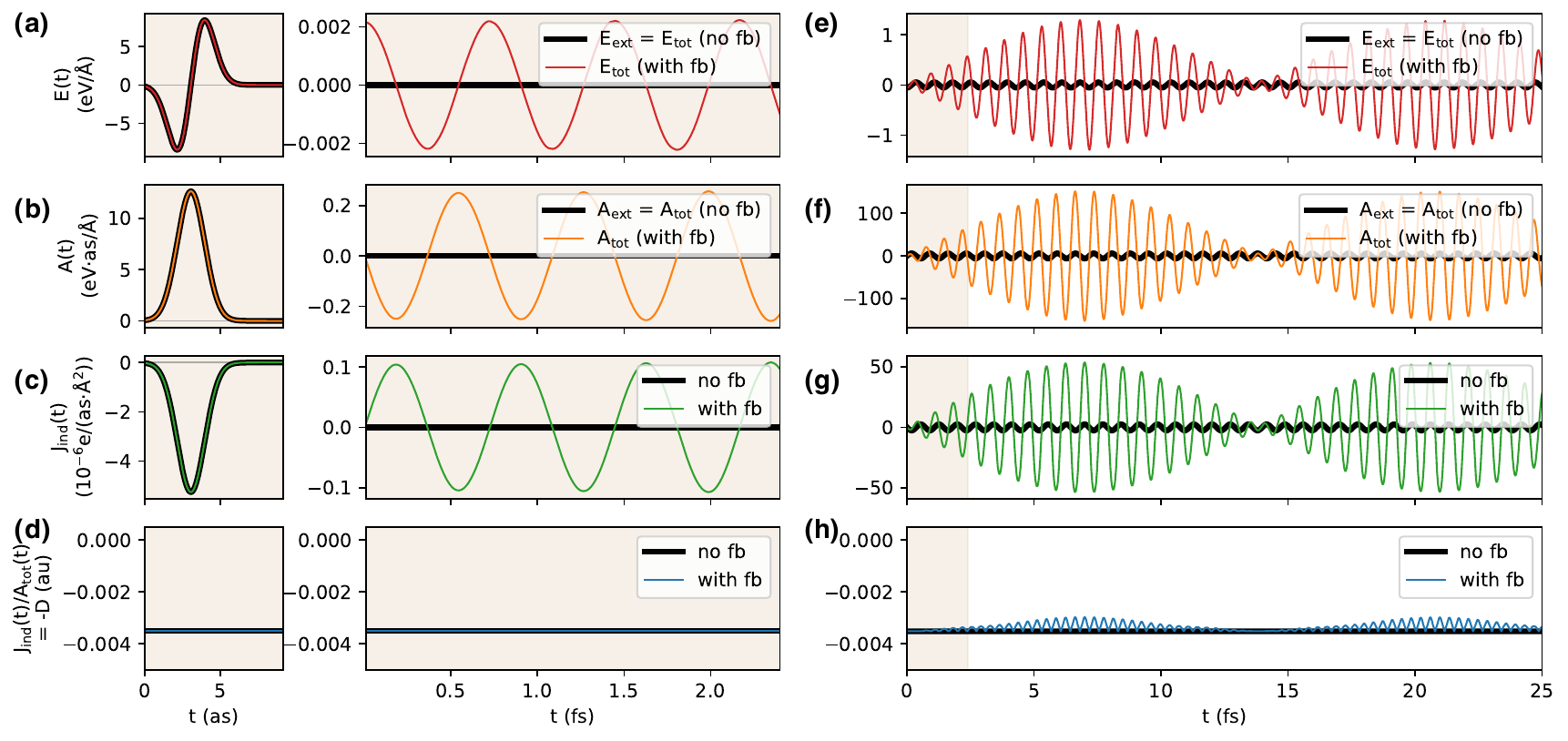}
    \caption{Noninteracting electrons. (a-d) Delta-like light driver. (e-f) Time-periodic light driver that has an $A_{\mathrm{ext}}$-amplitude half the peak amplitude of the delta-like driver. (a, e) Electric field of driver. (b, f) Vector potential of driver. (c, g) Induced current density. (d, h) Ratio of $A_{\mathrm{tot}}$ and $J_{\mathrm{ind}}$ in Hartree atomic units. Legend: fb means that $A_{\mathrm{ind}}$ is fed back into $A_{\mathrm{tot}}$.
    }
    \label{fig:nonint_ele}
\end{figure*}

This paper is organized as follows. In Sec.~\ref{sec:nonintele}, we introduce the concept of currons in the context of noninteracting electrons, illustrating how their emergence is analogous to conventional plasmonic excitations. Section~\ref{sec:int_ele_theory} describes the theoretical framework of TT-$GW$ that is used to describe interacting electrons under non-equilibrium conditions in weakly and moderately correlated systems. Sec.~\ref{sec:int_ele} extends the concept of currons to include the effects of el-el interactions.  Finally, in Sec.~\ref{sec:conclusion}, we summarize our findings and discuss potential experimental signatures of currons, as well as broader implications for optoelectronic and quantum materials research.

\section{Currons Generated by noninteracting Electrons} \label{sec:nonintele}
In order to better understand the characteristics of the induced current, we first discuss the creation of currons in the limit of one-electron theory, i.e., when interactions between the electrons (i.e., el-el interactions), $V$, are treated in a mean-field approach or simply turned off. This approximation dramatically simplifies the calculation of $\mathbf{J_{\mathrm{ind}}}(t)$ and $\mathbf{A_{\mathrm{tot}}}(t)$. 

Since we are interested in time evolution of lattice-periodic systems, we derive our equations in the momentum ($\mathbf{k}$) domain, where $\mathbf{k}$ is the crystal momentum in the first Brillouin zone. Starting from the effective low-energy tight-binding Wannier Hamiltonian of Na downfolded from the time-independent Kohn-Sham DFT Hamiltonian, $H^0_{\mathrm{TB}}(\mathbf{k})$ (see Appendix~\ref{sec_si:dft} for computational details on DFT, Appendix~\ref{sec_si:vx=0} for symmetry analysis and Appendix~\ref{sec_si:w90} for the construction of the Wannier Hamiltonian), we investigate the non-perturbative time-dependent effects from the light driver through the Peierls substitution, which implements minimal coupling in lattice systems~\cite{Peierls:1933aa, PhysRev.84.814, PhysRev.115.1460},
\begin{align} \label{eq:H_KE}
H^0_{\mathbf{k}} (t) = H^0_{\mathrm{TB}}\left(\mathbf{k} - \frac{q \mathbf{A_{\mathrm{tot}}}(t)}{\hbar}\right),
\end{align}
where $H^0_{\mathbf{k}} (t)$ is the time-dependent one-body Hamiltonian and  $q$ is the charge of an electron that carries a sign. 
Using $H^0_{\mathbf{k}} (t)$, we can obtain the velocity of the electrons,
\begin{align}
\mathbf{v}_{\mathbf{k}}(t) 
&= \frac{1}{\hbar} \mathbf{\nabla_{k}} H^0_{\mathbf{k}} (t) \\
&= \frac{1}{\hbar} \mathbf{\nabla_{k}} H^0_{\mathrm{TB}} \left(\mathbf{k}-
  \frac{q \mathbf{A_{\mathrm{tot}}}(t)  }{\hbar} \right), \label{eq:velocity_k_main}
\end{align}
from which the macroscopic current density can be calculated,
\begin{align} \label{eq:jtot_main}
\mathbf{J}_{\mathrm{tot}}(t)
&= \frac{q}{N_{\mathbf{k}}} \sum_{\mathbf{k}} n_{\mathbf{k}}(t) \mathbf{v}_{\mathbf{k}}(t),
\end{align}
where $N_{\mathbf{k}}$ is the number of $\mathbf{k}$-points, $n_{\mathbf{k}}$ is the number density of electrons at a $\mathbf{k}$-point. The induced current density is given by,
\begin{equation}
  \mathbf{J}_{\mathrm{ind}}(t) 
  = \mathbf{J}_{\mathrm{tot}}(t) - \mathbf{J}_{\mathrm{tot}}(t_0), \label{eq:jind=jtot_diff} \\
\end{equation}
where $t_0$ is the time at which the time-dependent external optical field is turned on, and  $\mathbf{J}_{\mathrm{ind}}(t) = \mathbf{J}_{\mathrm{tot}}(t)$ since $\mathbf{J}_{\mathrm{ext}}(t)=0$  and $\mathbf{J}_{\mathrm{tot}}(t<t_0)=0$ at equilibrium.

The induced current density $\mathbf{J}_{\mathrm{ind}}(t)$ generates an induced vector potential, $\mathbf{A}_{\mathrm{ind}}(t)$, as governed by the d'Alembert wave equation in the Coulomb gauge sourced by $\mathbf{J}_{\mathrm{ind}}(t)$~\cite{Brill1967},
\begin{align}
\epsilon_0 \frac{d^2 \mathbf{A}_{\mathrm{ind}}^{\perp}}{dt^2}
&= \mathbf{J}_{\mathrm{ind}}^{\perp}(t), \label{eq:dAlembert}
\end{align}
with $\mathbf{A}_{\mathrm{ind}}^{\parallel} = \mathbf{J}_{\mathrm{ind}}^{\parallel}$ = 0. Here, $\perp$ and $\parallel$, respectively, denote the transverse and longitudinal components of the field/current. Through Eqs.~\eqref{eq:velocity_k_main}-\eqref{eq:jind=jtot_diff}, we see that $\mathbf{J}_{\mathrm{ind}}^{\perp}(t)$ is a functional of $\mathbf{A}_{\mathrm{ind}}^{\perp}$, and since
\begin{equation}
\mathbf{A_{\mathrm{tot}}}(t) = \mathbf{A_{\mathrm{ext}}}(t) + \mathbf{A_{\mathrm{ind}}}(t), \label{eq:Atot=Aext+Aind_main}
\end{equation}
Eq.~\eqref{eq:dAlembert} can be rewritten as,
\begin{align}
\epsilon_0 \frac{d^2 \mathbf{A}_{\mathrm{ind}}^{\perp}}{dt^2} 
&= \mathbf{J}_{\mathrm{ind}}^{\perp}[\mathbf{A}_{\mathrm{ind}}^{\perp}+\mathbf{A}_{\mathrm{ext}}^{\perp} ]
\nonumber
\end{align}
which is a closed equation that we use to solve for $\mathbf{A}_{\mathrm{ind}}$ and $\mathbf{J}_{\mathrm{ind}}$ numerically.

Note that a time-dependent homogeneous $\mathbf{A}_{\mathrm{ext}}(t)$ has zero curl, so it produces no magnetic field ($\mathbf{B} = \boldsymbol{\nabla} \times \mathbf{A}=0$). 
In the special case of a homogeneous light field acting on a homogeneous electron gas, the scalar potential $\phi_{\mathrm{tot}}(t)$ is spatially constant, so its gradient vanishes. Faraday’s law therefore simplifies to
\begin{equation}
  \mathbf{E}_{\mathrm{tot}}(t) = - 
  \frac{d \mathbf{A}_{\mathrm{tot}}(t)}{dt}.\label{eq:Faraday's_law}
\end{equation}
Consequently, a homogeneous time-dependent $\mathbf{A}_{\text{ext}}(t)$
induces a homogeneous time-varying electric field even though no magnetic field accompanies it. 

The negative sign in Eq.~\eqref{eq:Faraday's_law} reflects the Lenz’s law: the induced electric field opposes the change in vector potential that created it, ensuring that the field resists changes in the current that sources it. This opposing field provides the restoring force, $\mathbf{f}_\mathrm{res}$, that pulls the charge displacement back toward zero. In Appendix~\ref{sec_si:restoring_force}, we derive the analytic expression of this restoring force (density) for a homogeneous, collisionless, noninteracting electron gas.

\subsection{Single-pulse driver} \label{sec:pulse_driver}
We begin by exciting the system using a single delta-like light pulse, since it can be considered as a superposition of light of all frequencies and we would like to probe the entire frequency space. The external pulse is modeled using a vector potential, $\mathbf{A}_{\mathrm{ext}}(t)$,  described by a Gaussian function that has a full width at half maximum (FWHM) of 2.1~atto seconds ($\si{as}$) and a peak amplitude of 12.6~$\si{eV\cdot as/\AA}$ (Fig.~\ref{fig:nonint_ele}b (left)), corresponding to a fluence of 4.1~$\si{mJ/cm^2}$. The corresponding external electric field is plotted in Fig.~\ref{fig:nonint_ele}a (left), which is related to the vector potential through the Faraday's law (Eq.~\eqref{eq:Faraday's_law}).
In this work, we use the units of \si{eV/ \AA} and \si{eV\cdot as / \AA} for the electric field, $E(t)$, and vector potential, $A(t)$, respectively, implicitly multiplying them with the elementary charge, since they are the natural choices at the atomistic scale. For $J_{\mathrm{ind}}$ and $A_{\mathrm{tot}}$, we will use the Hartree atomic unit (au) for it is more intuitive, as we will see.

First, we investigate a system with no feedback of the induced vector potential, $\mathbf{A_{\mathrm{ind}}}$, to the total vector potential, $\mathbf{A_{\mathrm{tot}}}$. As expected, since $\mathbf{A_{\mathrm{ext}}}(t)$ is polarized in the $x$-direction, the calculated induced current, $\mathbf{J_{\mathrm{ind}}}(t)$, is also polarized in the $x$-direction. Furthermore, in the absence of feedback from $\mathbf{A_{\mathrm{ind}}}(t)$, $\mathbf{J_{\mathrm{ind}}}(t)$ is an instantaneous response to the external driver, $\mathbf{A_{\mathrm{ext}}}(t)$ (Fig.~\ref{fig:nonint_ele}b-c). Also, $\mathbf{J_{\mathrm{ind}}}(t)$ ceases to exist the instant the external pulse ends and $\mathbf{J_{\mathrm{ind}}}(t)$ is directly proportional to $\mathbf{A_{\mathrm{ext}}}(t)$, indicating no phase shift or retardation effects (Fig.~\ref{fig:nonint_ele}d). Notably, the proportionality constant of $J_{\mathrm{ind}}/A_{\mathrm{tot}} =-0.0035$~\si{au} is negative, i.e., $\mathbf{J_{\mathrm{ind}}}(t)$ points in the opposite direction of $\mathbf{A_{\mathrm{ext}}}(t)$, indicating that the induced current opposes the change in the vector potential, consistent with the Lenz's Law. We will see in the next section that $J_{\mathrm{ind}}/A_{\mathrm{tot}}=-D$ is a material property that is, in fact, independent of the light driver.

Next, we consider the feedback effects of $\mathbf{A_{\mathrm{ind}}}$, such that the induced vector potential, $\mathbf{A_{\mathrm{ind}}}$, now contributes to the total vector potential, $\mathbf{A_{\mathrm{tot}}}$. Even though the contribution of $\mathbf{A_{\mathrm{ind}}}$  to $\mathbf{A_{\mathrm{tot}}}$ is negligible during the pulse due to their relative amplitudes (Fig.~\ref{fig:nonint_ele}a,b (left)), $\mathbf{A_{\mathrm{ind}}}$ noticeably persists after $\mathbf{A_{\mathrm{ext}}}(t)$ has been switched off, due to that $\mathbf{J_{\mathrm{ind}}}(t)$ continues to oscillate in a self-sustaining manner. Furthermore, $\mathbf{J_{\mathrm{ind}}}(t)$ continues to be proportional to the total vector potential, $\mathbf{A_{\mathrm{tot}}}(t)$, with the ratio of $J_{\mathrm{ind}}/A_{\mathrm{tot}}$ not only remaining negative, as in the absence of feedback (Fig.~\ref{fig:nonint_ele}b-d), also retaining the same value of $-0.0035$~\si{au}. 

Moreover, $\mathbf{J_{\mathrm{ind}}}(t)$ oscillates at the quantized frequency of $5.8$~\si{eV}, which falls within the experimentally accessible regime of ultraviolet light. Its amplitude of 0.11~\si{\mu e/(as\cdot \AA^2)} is significant and experimentally measurable. Comparing it to the drift current in Na, $\mathbf{J_{\mathrm{ind}}}(t)$ is orders of magnitude larger. A back-of-the-envelope calculation yields a drift current density of $1.2\times 10^{-3}$~\si{\mu e/ (as \cdot \AA^2)} in Na under a moderate $E$-field of $10^3$~\si{V/m} (given that its electrical conductivity is $2.0\times10^7$~\si{S/m}).

The analysis above marks the clear emergence of a quasiparticle, which we suggest to be named \textit{curron}. Given its frequency and amplitude we suggest that the \textit{curron} is experimentally detectable. The frequency of $5.8$~\si{eV}, which defines the curron of Na, is an intrinsic property of the host material. A curron is analogous to a plasmon, but instead of being a collective excitation of oscillating electron density, it is a collective excitation of oscillating currents (Fig.~\ref{fig:plasmon-vs-curron}). By allowing the induced vector potential generated by the current ($\mathbf{A_{\mathrm{ind}}}(t)$) to interact with the external vector potential ($\mathbf{A_{\mathrm{ext}}}(t)$), we have effectively constructed a system of \textit{interacting currents}, mediated by vector potentials, $\mathbf{A}$. Currons would not have existed in the absence of current-current interactions just as plasmon would not have existed in the absence of el-el interactions, as it is the el-el interactions that provide the restoring force for a plasmon.

Notably, a curron is not a plasmon, as the example above has clearly demonstrated, since el-el interactions have been explicitly turned off. This can be also implied from the continuity equation,
\begin{equation}
\frac{\partial \rho}{\partial t} + \boldsymbol{\nabla} \cdot \mathbf{J} = 0, \label{eq:continuity_eq}
\end{equation}
where $\rho(t)=qn(t)$ is the charge density. For a homogeneous current ($\boldsymbol{\nabla} \cdot \mathbf{J} = 0$), no charge density fluctuation is generated, meaning a purely homogeneous time-dependent $\mathbf{A}(t)$ cannot excite a plasmon. 

Similarly, a curron is also not an exciton. An exciton couples to the scalar  potential via its electric dipole moment and exists in materials that possess a bandgap allowing for bound electron-hole pairs. Here, neither the scalar potential nor the bandgap is present.

\subsection{Continuous time-periodic drive} \label{sec:continuous_driver}
With a single-pulse driver, $\mathbf{A_{\mathrm{ind}}}(t)$ (Fig.~\ref{fig:nonint_ele}b (right)) is perturbative: its peak value is always a fraction of that of $\mathbf{A_{\mathrm{ext}}}(t)$ (Fig.~\ref{fig:nonint_ele}b (left)). We next consider to nonperturbatively drive the system out of equilibrium using time-periodic light. By using a continuous light source, we supply the system with a constant source of energy. Furthermore, having identified the resonant frequency of the curron, we will fix the light source to this frequency (5.8~\si{eV}). We set the amplitude of $\mathbf{A}_\mathrm{ext}$ to be half the peak amplitude of the pulse in the previous section. 

As expected, the amplitudes of $\mathbf{E}_\mathrm{tot}$ and $\mathbf{A}_\mathrm{tot}$ become strongly amplified, reaching 23-24\,times the amplitudes of the external driver fields, $\mathbf{E}_\mathrm{ext}$ (Fig.~\ref{fig:nonint_ele}e) and $\mathbf{A}_\mathrm{ext}$ (Fig.~\ref{fig:nonint_ele}f). This large amplification confirms that the system is outside the linear-response regime. Through feedback, the induced vector potential $\mathbf{A}_\mathrm{ind}$ also boosts the induced current density $\mathbf{J}_\mathrm{ind}$, making it about 20\,times stronger than in the no-feedback case (Fig.~\ref{fig:nonint_ele}g). Most interestingly, although the external driver supplies a continuous wave at a single frequency, the total fields ($\mathbf{E}_\mathrm{tot}$ and $\mathbf{A}_\mathrm{tot}$) and the induced current ($\mathbf{J}_\mathrm{ind}$) display clear beating. Each quantity forms a wave packet whose envelope and carrier can be written as $\sin(\omega_\mathrm{env} t)\sin(\omega_\mathrm{car} t)$, with $\hbar\omega_\mathrm{env}=0.1$\,\si{eV} and $\hbar\omega_\mathrm{car}=5.7$\,\si{eV}. The envelope arises from the superposition of two components at $\hbar\omega_{1}=5.6$\,\si{eV} (softened transverse-current mode) and $\hbar\omega_{2}=5.8$\,\si{eV} (fixed laser frequency). The frequency splitting reflects the coexistence of two distinct modes: a homogeneous oscillation at the feedback-softened intrinsic frequency $\tilde\omega_c = 5.6$\,\si{eV}, and a driven response at the fixed laser frequency $\omega_d = 5.8$\,\si{eV}. The value of $\tilde\omega_c$ is determined solely by the steady-state amplitude of $\mathbf{A}_{\mathrm{tot}}$, and does not depend on the details of how the external field is turned on.

Furthermore, other than fluctuations about a baseline of $J_{\mathrm{ind}}/A_{\mathrm{tot}}= -D =-0.0035$~\si{au}, $D$ at resonance (Fig.~\ref{fig:nonint_ele}h) remain the same as that off-resonance, with and without feedback (Fig.~\ref{fig:nonint_ele}d). This suggests that $D$ is a material property that is a constant in the linear-response limit. To probe this deeper, we will now derive an analytical expression for $D$ for a simple model in the linear-response limit of perturbatively small $\mathbf{A}_{\mathrm{ext}}(t)$. 

\subsection{Linear response limit} \label{sec:linearresponselimit}
Substituting  Eq.~\eqref{eq:jtot_main} into Eq.~\eqref{eq:jind=jtot_diff}, $\mathbf{J}_{\mathrm{ind}}$ in the linear response limit can be expressed as,
\begin{align}
  \mathbf{J}_{\mathrm{ind}}(t) 
  &= \mathbf{J}_{\mathrm{tot}}(t) - \mathbf{J}_{\mathrm{tot}}(t_0) \nonumber\\ 
  &= \frac{1}{N_{\mathbf{k}}} 
     \sum_{\mathbf{k}} q\,n_{\mathbf{k}}(t)\,\mathbf{v}_{\mathbf{k}}(t) 
     - \frac{1}{N_{\mathbf{k}}} 
     \sum_{\mathbf{k}} q\,n_{\mathbf{k}}(t_0)\,\mathbf{v}_{\mathbf{k}}(t_0) \nonumber\\
  &= \frac{q}{\hbar}
     \frac{1}{N_{\mathbf{k}}} \sum_{\mathbf{k}} \bigg[
     n_{\mathbf{k}}(t)\,\boldsymbol{\nabla}_{\mathbf{k}} 
     H^0_{\mathrm{TB}} \bigg( \mathbf{k} - \frac{q\,\mathbf{A}_{\mathrm{tot}}(t)}{\hbar} \bigg) \nonumber\\
  &\quad\quad\quad\quad\quad\quad 
     - n_{\mathbf{k}}(t_0)\,\boldsymbol{\nabla}_{\mathbf{k}} H^0_{\mathrm{TB}}(\mathbf{k}) \bigg] \nonumber\\
  &= \frac{q^2}{\hbar^2}
     \frac{1}{N_{\mathbf{k}}} \sum_{\mathbf{k}} \bigg[
     \underbrace{[n_{\mathbf{k}}(t) - n_{\mathbf{k}} (t_0)] 
     \,\boldsymbol{\nabla}_{\mathbf{k}} H^0_{\mathrm{TB}} ( \mathbf{k})}_{\text{paramagnetic}} \nonumber\\
  &\quad\quad\quad\quad\quad\quad
   \underbrace{-n_{\mathbf{k}}\,\boldsymbol{\nabla}_{\mathbf{k}} 
     \big[ \boldsymbol{\nabla}_{\mathbf{k}} H^0_{\mathrm{TB}} (\mathbf{k}) 
     \cdot \mathbf{A}_{\mathrm{tot}}(t) \big]}_{\text{diamagnetic}} \bigg]
\end{align}
to linear order, where the first and second terms in the last line correspond to the paramagnetic and diamagnetic terms, respectively. In the velocity gauge that we are using, a homogeneous field preserves the canonical momentum $ \hbar \mathbf{k} $ in the absence of el-el interactions. As a result, $n_{\mathbf{k}}(t) = n_{\mathbf{k}} (t_0)$ and the paramagnetic term vanishes.  $\mathbf{J}_{\mathrm{ind}}(t)$ reduces to,
\begin{align}
\mathbf{J}_{\mathrm{ind}}(t) 
  &= - \frac{q^2}{\hbar^2}
    \frac{1}{N_{\mathbf{k}}} \sum_{\mathbf{k}} n_{\mathbf{k}}
    [ \mathbf{A}_{\mathrm{tot}}(t) \cdot \boldsymbol{\nabla}_{\mathbf{k}} ] \boldsymbol{\nabla}_{\mathbf{k}} H^0 (\mathbf{k})
    \label{eq:Jind_main} \\
  J_{\mathrm{ind}}^i(t) 
  &= - \bigg[ \frac{q^2}{\hbar^2}
    \frac{1}{N_{\mathbf{k}}} \sum_{\mathbf{k}} n_{\mathbf{k}}
    \frac{\partial^2 H^0_{\mathrm{TB}} (\mathbf{k})}{\partial k_i \partial k_j}  \bigg] {A}_{\mathrm{tot}}^j(t) \nonumber \\
  &= - D_{ij} {A}_{\mathrm{tot}}^j(t), \label{eq:Jvsalpha2A_main}
\end{align}
where repeated indices are summed. Here, we have introduced the material-dependent tensor, $D_{ij}$,
\begin{align}
  D_{ij}
  &= \frac{q^2}{\hbar^2}
    \frac{1}{N_{\mathbf{k}}} \sum_{\mathbf{k}} n_{\mathbf{k}}
    \frac{\partial^2 H^0_{\mathrm{TB}} (\mathbf{k})}{\partial k_i \partial k_j} , \label{eq:alpha_ij_main}
\end{align}
where $\frac{\partial^2 H^0_{\mathrm{TB}} (\mathbf{k})}{\partial k_i \partial k_j} $ is the second derivative of energy with respect to $k_i$ and $k_j$, representing the curvature of the energy band from the tight-binding Hamiltonian $H^0_{\mathrm{TB}}$ (or any similar Hamiltoniain representing the electronic structure).

Equation~\eqref{eq:Jvsalpha2A_main} defines a linear relationship between the induced current density $ \mathbf{J}_{\mathrm{ind}}(t) $ and the total vector potential $ \mathbf{A}_{\mathrm{tot}}(t) $. From this expression, it is evident that the proportionality constant $ J_{\mathrm{ind}}/A_{\mathrm{tot}} $ must be negative, which is a direct reflection of Lenz’s law, as demonstrated in Sections~\ref{sec:pulse_driver} and~\ref{sec:continuous_driver}. The tensor, $ D_{ij} $, quantifies the extent to which the current density responds to the applied electromagnetic field. 
In anisotropic materials, $ D_{ij} $ is a full tensor with possible off-diagonal components, reflecting the coupling between different spatial directions.

For isotropic systems with a parabolic energy dispersion given by $H^0 (\mathbf{k}) = \frac{\hbar^2 k^2}{2 m_{\mathrm{eff}}}$ where $m_{\mathrm{eff}}$ is the effective mass of the electron, its second derivative reduces to $\frac{\partial^2 H^0 (\mathbf{k})}{\partial k_i \partial k_j}  = \frac{\hbar^2}{m_{\mathrm{eff}}} \delta_{ij}$, simplifying $D_{ij}$ to
\begin{align}
  D_{ij}
  &= \frac{q^2}{\hbar^2}  \frac{\hbar^2}{m_{\mathrm{eff}}} \delta_{ij}
    \frac{1}{N_{\mathbf{k}}}  \sum_{\mathbf{k}} n_{\mathbf{k}} \nonumber \\
  &= \frac{q^2}{m_{\mathrm{eff}}} n_e \delta_{ij},
\end{align}
where $n_e = \frac{1}{N_{\mathbf{k}}} \sum_{\mathbf{k}} n_{\mathbf{k}}$ is the number density of electrons. The induced current density in Eq.~\eqref{eq:Jvsalpha2A_main} then simplifies to,
\begin{align}
  \mathbf{J}_{\mathrm{ind}}(t) 
  = - D {\mathbf{A}}_{\mathrm{tot}}(t), \label{eq:jind_main}
\end{align}
where 
\begin{equation}
D = \frac{q^2 n_e}{m_{\mathrm{eff}}}
\label{eq:alpha_def_main}
\end{equation}
takes the familiar form of the diamagnetic response in isotropic media. This 
confirms that $D$ is a material property that depends on only $q$, $n_e$ and $m_{\mathrm{eff}}$. Clearly, $D$ is independent of whatever that is driving the oscillation (e.g., the amplitude and frequency of the light). Since $D$ depends on $n_e$, it means that a lower electron density (e.g., obtained by lowering the chemical potential, $\mu$) will lead to $\mathbf{J}_{\mathrm{ind}}(t)$ of a smaller amplitude (Eq.~\eqref{eq:jind_main}). As we shall see later, this will also lead to a resonant current occurring at a lower frequency (Eq.~\eqref{eq:wc_main}).

We now verify our derivation by substituting in the values of $n_e$ and $m_{\mathrm{eff}}$ from our model. For Na, there is one conducting 3$s$ electron per unit cell, which has a volume of 266.5~\si{bohr}$^3$~\cite{Wyckoff1963}. Even though Na does not have a perfectly parabolic energy dispersion, its $m_{\mathrm{eff}}$ can be approximated as 0.97 $m_e$ by fitting the DFT 3$s$ band at the $\Gamma$-point to a parabola. This effectively approximates Na, a nearly free-electron system, as a homogeneous electron gas. Substituting these free-electron gas parameters into Eq.~\ref{eq:alpha_def_main} gives $D=0.0036$~\si{au}, in excellent agreement with the calculated results in Figs.~\ref{fig:nonint_ele}c, g.

Next, we relate \( D \) to the curron frequency, \( \omega_c \), by introducing the transverse dielectric function, \( \epsilon_T(\omega) \), which relates the external and total vector potentials via
\begin{equation} \label{eq:epsilon_T_def}
\mathbf{A}_{\mathrm{tot}}(t) = \epsilon_T(\omega)^{-1} \mathbf{A}_{\mathrm{ext}}(t).
\end{equation}
For simplicity, we assume that both \( \mathbf{A}_{\mathrm{tot}} \) and \( \mathbf{J}_{\mathrm{ind}} \) take the sinusoidal forms \( \mathbf{A}_{\mathrm{tot}}(t) = \mathbf{A}_0 e^{-i\omega t} \) and \( \mathbf{J}_{\mathrm{ind}}(t) = \mathbf{J}_0 e^{-i\omega t} \).

To derive \( \epsilon_T(\omega) \), we substitute Eqs.~\eqref{eq:Atot=Aext+Aind_main}, \eqref{eq:epsilon_T_def}, and \eqref{eq:alpha_def_main} into the d'Alembert wave equation (Eq.~\eqref{eq:dAlembert}),
\begin{align}
\epsilon_0 \frac{\partial^2} {\partial t^2} 
\{[1 - \epsilon_T(\omega)]
\mathbf{A}_{\mathrm{tot}}^{\perp}(t)\}
&= - D {\mathbf{A}}_{\mathrm{tot}}^{\perp}(t)
\nonumber \\
\epsilon_0 \frac{\partial^2} {\partial t^2} \mathbf{A}_{\mathrm{tot}}^{\perp}(t)
&= - \frac{ D }{1 - \epsilon_T(\omega) } {\mathbf{A}}_{\mathrm{tot}}^{\perp}(t)  \nonumber \\
-\omega^2 \epsilon_0 {\mathbf{A}}_{\mathrm{tot}}^{\perp}(t) 
&= - \frac{  D }{1 - \epsilon_T(\omega) } {\mathbf{A}}_{\mathrm{tot}}^{\perp}(t)  \nonumber \\
\epsilon_T(\omega) 
  &= 1 - \frac{\omega_c^2}{\omega^2} \label{eq:epsilon_T}
\end{align}
with
\begin{equation}
\omega_c^2 
= \frac{D} {\epsilon_0} 
= -\frac{J_0} {\epsilon_0 A_0} \label{eq:curron_freq_nonint}
\end{equation}
being the (resonant) curron frequency.
With $\omega_c=5.8~\si{eV}=0.21~\si{Ha}$, our calculations in Sec.~\ref{sec:pulse_driver} confirm this derivation. We further confirm that $\epsilon_T(\omega) <  1$. Above the resonant frequency, $0 < \epsilon_T(\omega) < 1$ and $\mathbf{A}_\mathrm{tot} \approx  \mathbf{A}_\mathrm{ext}$. Below the resonant frequency, $\epsilon_T(\omega) < 0$. This means that $\mathbf{A}_\mathrm{ind} \propto - \mathbf{A}_\mathrm{ext}$, leading to an attenuation of $\mathbf{A}_\mathrm{tot}$ (and a frequency blueshift, see Eq.~\eqref{eq:wc_main} below). Nonetheless, we would like to point out that even though $\mathbf{A}_\mathrm{tot}$ may be attenuated compared to $\mathbf{A}_\mathrm{ext}$, the higher frequency of $\mathbf{A}_\mathrm{tot}$ may lead to an $\mathbf{E}_\mathrm{tot}(t)$
that is not necessarily attenuated compared to $\mathbf{E}_\mathrm{ext}(t)$, according to the Faraday's Law (Eq.~\eqref{eq:Faraday's_law}). In an experiment, it is $\mathbf{E}_\mathrm{tot}(t)$ that is measured, not $\mathbf{A}_\mathrm{tot}$.

In fact, Eq.~\eqref{eq:epsilon_T} is reminiscent of the plasmon frequency, $\omega_p$, which is given by $\epsilon_L(\omega) = 1 - \omega_p^2 / \omega^2$, in the long-wavelength limit of $qc/\omega \to 0$, where $\epsilon_L(\omega)$ is the longitudinal dielectric function and $c$ is the speed of light. If we express Eq.~\eqref{eq:curron_freq_nonint} in terms of Eq.~\eqref{eq:alpha_def_main} in this limit, we get, 
\begin{equation}
 \omega_c = \sqrt{\frac{n_e q^2}{\epsilon_0 m_\mathrm{eff}}} \label{eq:wc_main}
\end{equation}
which turns out to be the plasmon frequency of a metal modeled with a parabolic dispersion. In other words $\omega_c=\omega_p$, in the limit of $qc/\omega \to 0$, which should be expected~\cite{Nozieres1959} since $\epsilon_T(\omega) = \epsilon_L(\omega)$ in this limit. 

Some readers will recognize $D$ in Eq.~\eqref{eq:alpha_def_main} as the Drude weight for free carriers in the standard Drude theory. It is often expressed as $D = \epsilon_0 \omega_p^2$ to highlight its relation to plasmons. The Drude weight, $D$, is the coefficient of the zero-frequency Dirac delta in the real part of the optical conductivity, $\operatorname{Re}\sigma(\omega)=\pi D\,\delta(\omega)+\sigma_{\mathrm{reg}}(\omega)$, where the $\delta(\omega)$ peak at $\omega = 0$ reflects the instantaneous, nondissipative diamagnetic response of the electron gas to the vector potential and $\sigma_{\mathrm{reg}}(\omega)$ being the regular non-singular component that describes dissipative absorption at $\omega>0$. The Drude weight is positive semi-definite ($D\ge 0$) and serves as an indicator of whether the material can host coherent current-density oscillations. If the host is metallic or superconducting, then $D>0$; the system supports dissipationless currents, and the transverse current resonance associated with the curron can emerge. In contrast, $D=0$ in an insulator; the vector potential cannot induce a persistent bulk current, and no curron resonance occurs. Instead, the resonant mode corresponds to the oscillation of dipole moment~\cite{Crisp1969}.

\section{Theory of Interacting Electrons Out of  Equilibrium} \label{sec:int_ele_theory}

Having established the curron as an experimentally observable quasiparticle in the noninteracting-electron limit, we now investigate how el-el interactions affect its spectral properties, e.g., its frequency and amplitude. To this end, we employ a state-of-the-art time-dependent TT-$GW$ formalism that rigorously incorporates many-body interactions beyond the mean-field approximation. 

Within Hedin's $GW$ formalism~\cite{Hedin1965}, we solve the Kadanoff-Baym equations~\cite{Stefanucci:2013oq} via explicit time-stepping on the real-time axis to calculate non-equilibrium electron dynamics. The single-particle Green's function $G_{\mathbf{k}}(t, t') = -i \langle \mathcal{T}_\mathcal{C} c_{\sigma \mathbf{k}}(t)\, c^\dagger_{\sigma \mathbf{k}}(t') \rangle$ is computed on the Keldysh contour $\mathcal{C}$, which comprises the forward and backward real-time branches as well as a vertical imaginary-time track that encodes initial thermal correlations. The $GW$ self-energy $\Sigma_{\mathbf{k}}(t, t')$, which resums all electronic polarization bubbles within the random-phase approximation, is evaluated self-consistently along the same contour during the time evolution.

In practice, the interacting Green's function, $G_{\mathbf{k}}(t,t')$, needed to calculate $\Sigma_\mathbf{k}(t,t')$ is not known \textit{a priori}; for every time step, $t$, $G_{\mathbf{k}}(t,t')$ has to be calculated   self-consistently starting from an initial guess. In the first iteration of each self-consistent $GW$-cycle, $G_{\mathbf{k}}(t,t')$ is approximated as the noninteracting Green's function, $G^0_{\mathbf{k}}(t,t')$, calculated from the same effective low-energy tight-binding Hamiltonian, $H^0_{\mathbf{k}} (t)$, used in  Sec.~\ref{sec:nonintele} for noninteracting electrons and defined in Eq.~\eqref{eq:H_KE}. With this guess of $G_{\mathbf{k}}(t,t')$, $\Sigma^{\mathrm{GW}}_{\mathbf{k}}(t, t')$ is calculated in the following order,
\begin{align}
  P_{\mathbf{q}}(t, t') &=
  - \frac{i\hbar}{N_{\mathbf{k}}} \sum_{\mathbf{k}}
  G_{\mathbf{k} + \mathbf{q}}(t', t)
  G_{\mathbf{k}}(t, t'), \label{eq:P_q_main} \\
W_{\mathbf{q}}(t, t') &= V_{\mathbf{q}}
  + V_{\mathbf{q}} \cdot [ P_{\mathbf{q}} \ast W_{\mathbf{q}} ] (t, t'), \label{eq:W_q_main} \\
\Sigma^{\mathrm{GW}}_{\mathbf{k}}(t, t') &=
  \frac{i\hbar}{N_\mathbf{q}} \sum_{\mathbf{q}}
  G_{\mathbf{k} - \mathbf{q}}(t, t')
  W_{\mathbf{q}} (t, t'),  \label{eq:S_q_main}
\end{align}
where $P_{\mathbf{q}}$ is the noninteracting polarizability, $V_{\mathbf{q}}$ is the bare electron interaction (see Appendix \ref{sec_si:w90}), $W_{\mathbf{q}}$ the screened Coulomb interaction, and $\ast$ denoting a convolution over the complex-time contour, $\mathcal{C}$. Having obtained $\Sigma^{\mathrm{GW}}_{\mathbf{k}}(t, t')$, we solve the Dyson equation in the integro-differential form, also known as the Kadanoff-Baym equation,
\begin{equation}
[i \hbar \partial_t - H^0_{\mathbf{k}} (t)] G_{\mathbf{k}}(t,t') 
- \int_{\mathcal{C}} d \bar t \Sigma^{\mathrm{GW}}_{\mathbf{k}}(t,\bar t) G_{\mathbf{k}}(\bar t,t') = \delta_{\mathcal{C}}(t,t'), \label{eq:KBeq_main}
\end{equation}
for an updated $G_{\mathbf{k}}(t,t')$,

In the next and all subsequent iterations, $H^0_{\mathbf{k}}(t)$ is calculated and updated within the Green's function framework and $\Sigma^{\mathrm{GW}}_{\mathbf{k}}(t, t')$ is again calculated using Eqs.~\eqref{eq:P_q_main}, \eqref{eq:W_q_main} and \eqref{eq:S_q_main}. The $GW$-cycle
continues until $G_{\mathbf{k}}(t, t')$, $W_{\mathbf{q}}(t, t')$, $\Sigma^{\mathrm{GW}}_{\mathbf{k}}(t, t')$ and $P_{\mathbf{q}}(t, t')$ are all converged. In our implementation, we use the NESSi~\cite{Schuler2020} library to manage the non-equilibrium Green's functions.

For each iteration of a self-consistent $GW$-cycle, $H^0_{\mathbf{k}}(t)$ and therefore, $\mathbf{A_{\mathrm{tot}}}(t)$, have to be determined in what we call the $\mathbf{A_{\mathrm{tot}}}$-cycle. We first approximate $\mathbf{A_{\mathrm{tot}}}(t)$ to be $\mathbf{A_{\mathrm{ext}}}(t)$. With the approximated time-dependent Hamiltonian obtained using Eq.~\eqref{eq:H_KE}, the velocity of the electron is calculated using the semiclassical formula of Eq.~\eqref{eq:velocity_k_main}.
The total macroscopic current density $\mathbf{J}_{\mathrm{tot}}(t)$ due to the moving electron induced by light is then calculated using Eq.~\eqref{eq:jind=jtot_diff}. By summing over all $\mathbf{k}$-points, the microscopic spatial variations in the current are averaged out, and $\mathbf{J}_{\mathrm{tot}}(t)$ is homogeneous. The induced current, given by Eq.~\eqref{eq:jind=jtot_diff}, induces a homogeneous vector potential $\mathbf{A}_{\mathrm{ind}}(t)$, which is calculated using Eq.~\eqref{eq:dAlembert}.
With $\mathbf{A}_{\mathrm{ind}}(t)$, the value of $\mathbf{A}_{\mathrm{tot}}(t)$ is updated in the next iteration using Eq.~\eqref{eq:Atot=Aext+Aind_main}. This vector potential is also homogeneous. The self-consistent $\mathbf{A_{\mathrm{tot}}}$-cycle of solving Eqs.~\eqref{eq:H_KE}, \eqref{eq:velocity_k_main}, \eqref{eq:jtot_main}, \eqref{eq:jind=jtot_diff}, \eqref{eq:dAlembert} and \eqref{eq:Atot=Aext+Aind_main} is repeated until $\mathbf{A_{\mathrm{tot}}}(t)$ and $H^0_{\mathbf{k}}(t)$ are converged for each iteration of the self-consistent $GW$-cycle.

\begin{figure} [b] 
    \centering
    \includegraphics[width=\linewidth]{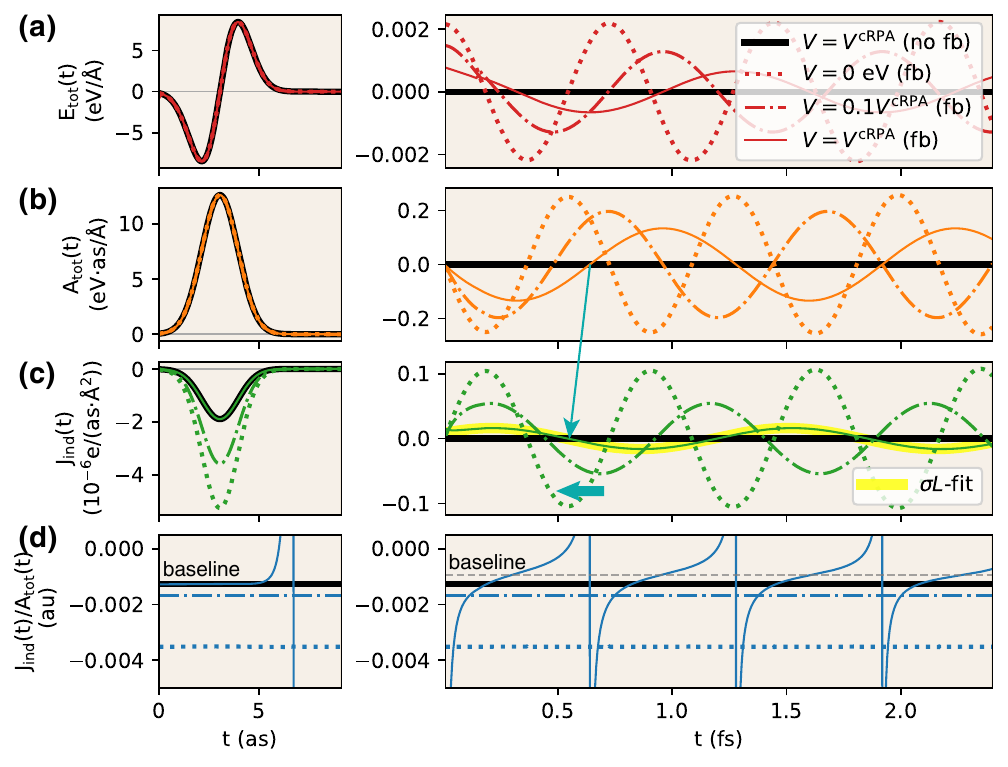}
    \caption{Interacting electrons driven by a delta-like light pulse. 
    (a) Total electric field of driver. 
    (b) Total vector potential of driver. 
    (c) Induced current density. The yellow line shows the $\sigma L$ fitted current density obtained from our RLC-model. The teal arrow shows the $32^\circ$ phase lead of $J_{\mathrm{ind}}$ over $A_{\mathrm{tot}}$ 
    (d) Ratio of $J_{\mathrm{ind}}$ to $A_{\mathrm{tot}}$ in Hartree atomic units. The gray dashed line on the right marks the elevated baseline, which corresponds to the baseline average in that region. Legend in (a, right) applies to all panels; fb means that $A_{\mathrm{ind}}$ is fed back into $A_{\mathrm{tot}}$. $V^{\mathrm{cRPA}}$ refers to the screened interaction computed within cRPA, as defined in Appendix~\ref{sec_si:w90}. Solid color line plots of Fig.~\ref{fig:nonint_ele}(a-d) are reproduced here in (a-d) as dotted lines}
    \label{fig:int_ele}
\end{figure}

\section{Currons Generated by Interacting electrons} \label{sec:int_ele}

To include the interaction between the electrons, we solve the Kadanoff-Baym equation numerically, focusing our attention on the single-pulse light driver. Our approach does not require the time dependence of the vector potential to be perturbative, only that the many-body interactions be perturbative relative to the kinetic energy of the system. As such, it is, in principle, beyond the linear response limit approximation of Sec.~\ref{sec:linearresponselimit}. This approach would allow us to capture non-perturbative time-dependent effects with accurate treatment of el-el interactions. 

For comparison, the results for noninteracting electrons with $\mathbf{A}_{\mathrm{ind}}(t)$-feedback (fb) of Fig.~\ref{fig:nonint_ele} (a-d, colored solid lines) are reproduced in Fig.~\ref{fig:int_ele} as colored dotted lines. This is superposed with the results for interacting electrons with $\mathbf{A}_{\mathrm{ind}}(t)$-feedback. One may note from Fig.~\ref{fig:int_ele} that the electric field due to el-el interactions significantly renormalizes the magnitude of $\mathbf{J}_{\mathrm{ind}}(t)$ (Fig.~\ref{fig:int_ele}c (left)) (by 2.8 times during the pulse and 6.6 times after the pulse). Recall that the electric field induced by a homogenous time-dependent $\mathbf{A}_{\mathrm{ind}}(t)$ through Eq.~\eqref{eq:dAlembert} hardly changes $\mathbf{J}_{\mathrm{ind}}(t)$ (Fig.~\ref{fig:nonint_ele}c). This should be expected since the electric fields from the electrons are orders of magnitude larger than the electric field of light. Moreover, the frequency of the curron is reduced 1.8 times from the UV regime to the visible light regime (3.2~\si{eV}). This should be expected as the el-el interactions increase the effective mass of the electron, increasing its inertia of motion. 

\subsection{Transverse vs longitudinal excitations}
Nonetheless, despite the strength of the electric field arising from el-el interactions, its effect on $\mathbf{A}_{\mathrm{ind}}(t)$ remains negligible relative to $\mathbf{A}_{\mathrm{ext}}(t)$ throughout the duration of the pulse (Fig.~\ref{fig:int_ele}b (left)). Moreover, these interactions are incapable of exciting a curron. When the restoring force provided by the oscillating electric field of light is switched off (by eliminating the feedback from $\mathbf{A}_{\mathrm{ind}}(t)$), both $\mathbf{A}_{\mathrm{ind}}(t)$ and $\mathbf{J}_{\mathrm{ind}}(t)$ immediately vanish once $\mathbf{A}_{\mathrm{ext}}(t)$ ceases (Fig.~\ref{fig:int_ele} (right)). This confirms that the current-density oscillation is driven by the electric field induced by vector field ($\mathbf{E}_{\mathrm{tot}}=-\frac{d\mathbf{A}_{\mathrm{tot}}}{dt}$) and not the electric field arising from el-el interactions ($\mathbf{E}_{\mathrm{tot}}=\boldsymbol{\nabla} \phi_{\mathrm{tot}}$).

This distinction highlights the fundamental difference between the two collective modes. While both currons and plasmons represent collective electron oscillations at characteristic frequencies of the electron density, they are inherently different in nature: a curron corresponds to a transverse oscillation of the current density, driven by a divergence-free field, whereas a plasmon involves a longitudinal oscillation of charge, driven by a field with non-zero divergence.

In this work, the external vector potential $\mathbf{A}_{\mathrm{ext}}(t)$ is homogeneous and time-dependent. It generates a purely transverse electric field through $\mathbf{E}_{\mathrm{ext}}(t) = -\frac {d\mathbf{A}_\mathrm{ext}}{dt}$, for which $\nabla \cdot \mathbf{E}_{\mathrm{ext}} = 0$ everywhere in the bulk. Consequently, the associated polarization field $\mathbf{P}(t) = \epsilon_0 \chi_e \mathbf{E}_{\mathrm{ext}}(t)$ is also divergence-free, leading to zero bound charge density, $\rho_{\mathrm{b}} = -\nabla \cdot \mathbf{P} = 0$. Gauss’s law then requires the total displacement field to satisfy $\nabla \cdot \mathbf{D}(t) = 0$, implying that the free charge density must also vanish, $\rho_{\mathrm{free}} = 0$. Under these conditions, longitudinal plasmons cannot be excited: there is no driving field to sustain a charge oscillation. However, a transverse mode like the curron can still be coherently excited, since it depends only on the dynamics of the current density and the feedback from transverse electric field induced by vector field $\mathbf{A}_{\mathrm{tot}}(t)$ driving it.

\subsection{Emergence of paramagnetic current} \label{sec:paramagnetic_cur}
Furthermore, $\mathbf{J}_{\mathrm{ind}}(t)$ now leads $-\mathbf{A}_{\mathrm{tot}}(t)$ by a phase of $32^{\circ}$, indicating that el-el interactions give rise to a retarded paramagnetic response in addition to the instantaneous diamagnetic term. The total induced current takes the form:
\begin{align}
\mathbf{J}_{\mathrm{ind}}(t)
  &= - \int_{t_0}^{t} dt'\;
       \boldsymbol{K}^{R}_{T}(t, t')\,
       \mathbf{A}_{\mathrm{tot}} \\
  &= \underbrace{-\,D\,\mathbf{A}_{\mathrm{tot}}(t)}_{\text{diamagnetic}}
  \underbrace{\;-\;\int_{t_0}^{t} dt'\;
              \boldsymbol{\chi}^{R}_{T}(t, t')\,
              \mathbf{A}_{\mathrm{tot}}(t')}_{\text{paramagnetic}},
\label{eq:Jind_retarded}
\end{align}
where $\boldsymbol{K}^{R}_{T}(t, t')$ is the kernel of the linear response, encapsulating both instantaneous and retarded effects. Here, the retarded part is expressed through the transverse current-current susceptibility $\boldsymbol{\chi}^{R}_{T}(t, t')$, which arises from el-el interactions. The emergence of the paramagnetic term is because $\Sigma^{\mathrm{GW}}_{\mathbf{k}}(t,t')$ breaks the conservation of $\mathbf{k}$ (Eq.~\eqref{eq:S_q_main}) in the velocity gauge. It captures the time-nonlocal response due to interaction-induced scattering and memory effects. Compared to the noninteracting case (Eq.~\eqref{eq:jind_main}, dotted line in Fig.~\ref{fig:int_ele}d), the diamagnetic response coefficient $D$ is reduced when el-el interactions are present (solid line, same figure) due to the larger $GW$-renormalized $m_\mathrm{eff}$ (see Eq.~\eqref{eq:alpha_def_main}). It also reflects the redistribution of spectral weight from the instantaneous diamagnetic response to the retarded paramagnetic response, consistent with the $f$-sum rule.

In Fig.~\ref{fig:int_ele}d, before the pump is applied (left), the solid colored line sits above the collision-free (dotted) Drude line due to equilibrium $GW$-corrections that renormalize the effective mass and reduce the diamagnetic weight from $D$ to $\tilde D$. When the strong pump field is switched on (right), the system enters a nonequilibrium steady state, causing the ratio $J_{\mathrm{ind}}/A_{\mathrm{tot}}$ shifts further upward to a new, elevated baseline that is less negative. This shift is driven by two main effects: (i) the strong self-consistent vector potential further softens the intrinsic mode frequency, slightly reducing $\lvert\tilde D\rvert$, and (ii) el-el interactions induce a retarded paramagnetic response described by the memory kernel $\boldsymbol{\chi}^{R}_{T}(t,t')$, generating a leading paramagnetic current component.Together, these effects raise the average value from $-\tilde D$ to the less negative level observed on the right.

This elevation of the baseline provides a direct signature of broken time-translation invariance and illustrates the emergence of genuine two-time physics captured by the Kadanoff-Baym formalism. A time-dependent external vector potential breaks the temporal symmetry of the system, making the induced current at time $t$ dependent on the entire prior history of the applied field through $\boldsymbol{\chi}^{R}_{T}(t,t')$. In contrast, any one-time $GW$ approach, whether formulated with equal-time adiabatic Green’s functions or instantaneous density matrices, can reproduce the static quasiparticle mass renormalization (and thus the left baseline) but, by construction, cannot generate the memory-driven paramagnetic current responsible for the additional upward shift to the right baseline. To our knowledge, this upward shift between the two baselines represents the first direct numerical observation of a two-time many-body effect in transverse electrodynamics. We propose that this predicted phenomenon can be experimentally verified using pump-probe spectroscopy, by measuring the transient ratio $J_{\mathrm{ind}}/A_{\mathrm{tot}}$ immediately following an intense optical excitation.

\subsection{Nonlocal $GW$ Quasiparticle renormalization} \label{sec:gw_renorm}

\begin{figure*} [!htbp]
    \centering
    \includegraphics[width=\linewidth]{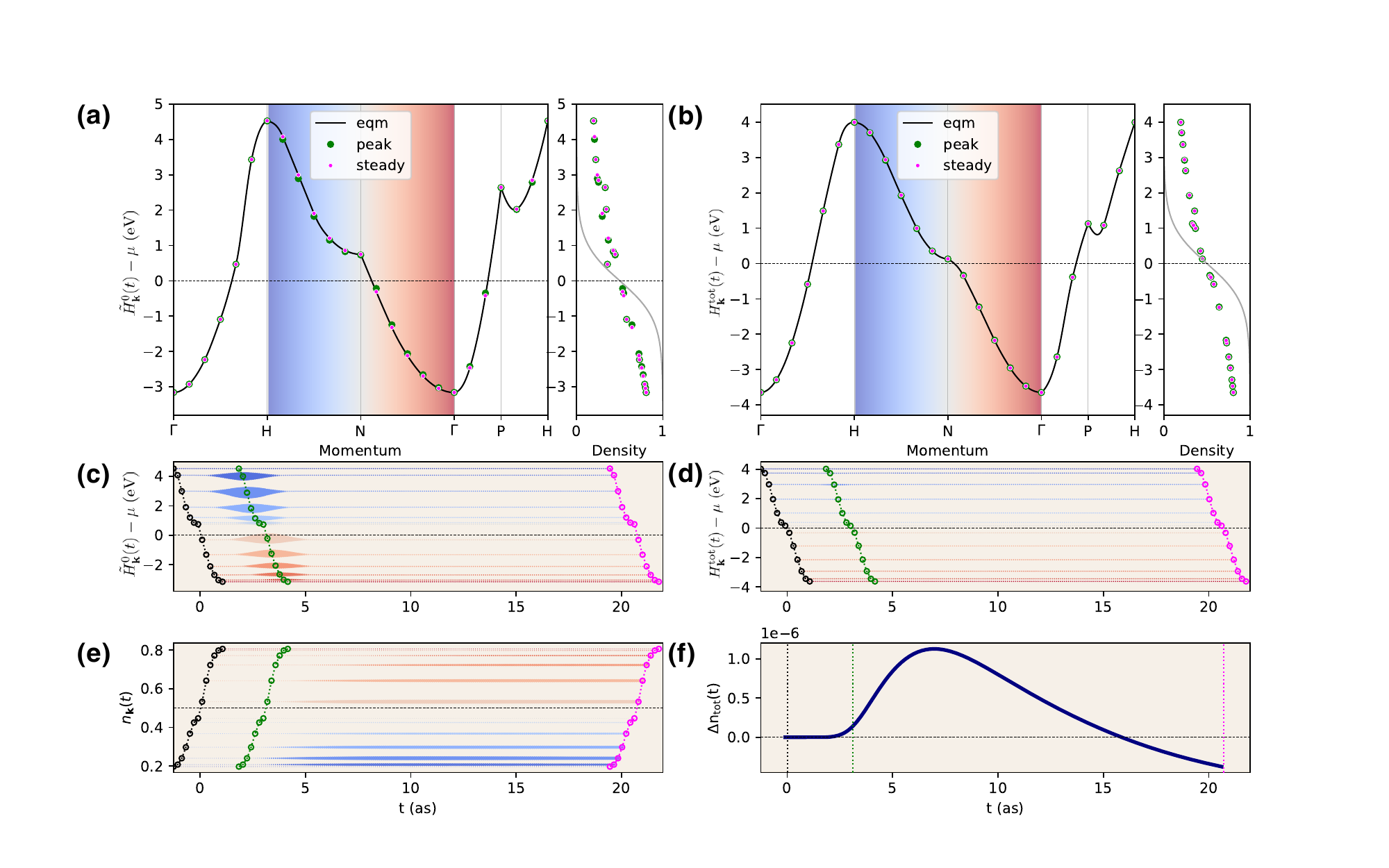}
    \caption{TT-$GW$ results. (a) $\tilde H^0_{\mathbf{k}}(t)$ relative to the chemical potential, $\mu$. (b) $H^\mathrm{tot}_{\mathbf{k}}(t)$ relative to $\mu$. (a, b) The black lines show interpolated energy at equilibrium, whereas the green and pink dots denote the energy at each $\mathbf{k}$-point when the pulse is at its peak amplitude (i.e., $t=t_p$) and at steady state (i.e., $t=t_s=20~\si{as}$), respectively. (c), (d) and (e) plot $\tilde H^0_{\mathbf{k}}(t)- \mu$, $H^\mathrm{tot}_{\mathbf{k}}(t)- \mu$ and $n_\mathbf{k}$, respectively, as a function of time, $t$, along the selected $\mathbf{k}$-path of $\mathrm{H} \rightarrow \mathrm{N} \rightarrow \mathrm{\Gamma}$ (and highlighted in (a) and (b) using a blue-white-red color map). Here, $n_\mathbf{k}(t)$ is electron density per primitive unit cell of Na. Each $\mathbf{k}$ point is time-shifted to reproduce the shape of the bandstructure and black, green and magenta hollow dots are used to mark $t=t_0$, $t_p$ and $t_s$, respectively. The area of each dot is proportional to the change of its magnitude relative to equilibrium. The largest dot size in (c) and (d) correspond to 0.1~\si{eV}, while the largest dot size in (e) corresponds to $6\times10^{-5}$. (f) Covergence study of the conservation of $n_\mathrm{tot}(t)$, where $n_\mathrm{tot}(t) = \frac{1}{N_{\mathbf{k}}} \sum n_\mathbf{k}(t)$.}
    \label{fig:nessi}
\end{figure*}

In order to understand how the el-el interactions resist change in $\mathbf{J}_{\mathrm{ind}}(t)$ due to light, we will separate the Hamiltonian, $H^{\mathrm{tot}}_{\mathbf{k}}(t)$, into the one-body part, $\tilde H^0_{\mathbf{k}} (t)$, and the many-body correction part, $H^\mathrm{corr}_{\mathbf{k}}(t)$. Here, $\tilde H^0_{\mathbf{k}} (t) = H^0_{\mathbf{k}} (t) - V^{\mathrm{xc}}_{\mathbf{k}}$, where $H^0_{\mathbf{k}} (t)$ is the minimally coupled DFT Hamiltonian as defined in Eq.~\eqref{eq:H_KE}, $V^{\mathrm{xc}}_{\mathbf{k}}$ is the DFT exchange-correlation potential and $H^\mathrm{corr}_{\mathbf{k}}(t) = \frac{1}{2} \operatorname{Im} \operatorname{Tr} [\Sigma^{\mathrm{GW}}_{\mathbf{k}} \ast G_{\mathbf{k}}]^<(t,t)$ is the Galitskii-Migdal energy (see Appendix~\ref{sec_si:tdgw} for details). In Figs.~\ref{fig:nessi}a, b, we plot the eigenvalues of our single-orbital Hamiltonian at $t=t_0$ (in black line), $t_\mathrm{p}$ (green dots), $t_\mathrm{s}$ (magenta dots), where $t=t_0$ stands for time at equilibrium (before the pulse), $t=t_p$ stands for time when the pulse is at its peak amplitude and $t=t_s=20~\si{as}$ stands for time at steady state after the pulse is over. 

The eigenvalues of the one-body part of the Hamiltonian, $\tilde H^0_{\mathbf{k}} (t)$, are plotted in Fig.~\ref{fig:nessi}a, along a special $\mathbf{k}$-path: $\Gamma \rightarrow 
H \rightarrow N \rightarrow \Gamma \rightarrow P \rightarrow 
H$ (see Appendix~\ref{sec_si:dft} for details). Symmetry constraints along the high-symmetry path leads to a symmetry-enforced $x$-component of the electron velocity, $v_x=0$ (see Eq.~\eqref{eq:velocity_k_main}), at special $\mathbf{k}$-points along the special $\mathbf{k}$-path. In particular, $v_x$ must vanish at the special points $\Gamma$, $H$, and $N$, and everywhere along the $\Delta$ line from $\Gamma$ to $H$ (see Appendix~\ref{sec_si:vx=0} for symmetry analysis). 

The interaction with an $x$-polarized field enters the one-body Hamiltonian through minimal coupling, $H^0_{\mathbf{k}} (t) = H^0_{\mathrm{TB}}(\mathbf{k} - \frac{q \mathbf{A_{\mathrm{tot}}}(t)}{\hbar})$ (Eq.~\eqref{eq:H_KE}), which to first order in $\mathbf{A}_{\mathrm{tot}}$, yields the following time-dependent perturbation, $H_{\mathrm{int}}(t)$,
\begin{equation} \label{eq:min_coupling_first_order}
H_{\mathrm{int}}(t) = - \frac{q A_{\mathrm{tot}}^{x}(t)}{\hbar} v_x.
\end{equation}
Along the segment $\Gamma \rightarrow H$ ($\Gamma$ and $H$, inclusive), which lies on the $\Delta$-line, symmetry enforces $v_x = 0$ throughout the entire segment, causing $H_{\mathrm{int}}$ to vanish identically. No intraband current can be generated by an $x$-polarized field along this direction to first order in the linear regime. This is evident from Fig.~\ref{fig:nessi}a, in which we see the green and magenta dots lying on top of each other on the black line for the entire segment. This shows that our choice of $\mathbf{A_{\mathrm{tot}}}(t)$ remains in the linear regime. A field with sufficiently large amplitude may shift the instantaneous momentum away from the $\Delta$-line, leading to finite $v_x$, but that is outside the scope of this work.

Along all other segments of the path, $H \rightarrow  N \rightarrow  \Gamma \rightarrow  P \rightarrow H$ (excluding $H$, $N$ and $\Gamma$), the symmetry of \textit{Im$\bar{3}$m} does not constrain $v_x$, which remains free to vary according to the band dispersion. In Fig.~\ref{fig:nessi}a, we see that the displacement of the green dot (representing peak $\mathbf{A}_{\mathrm{ext}}$ amplitude) is proportional to the gradient and carrying the same sign (in accordance to Eq.~\eqref{eq:min_coupling_first_order}). In Fig.~\ref{fig:nessi}c, we plot the magnitude of $\Delta H^0_{\mathbf{k}}(t) = \Delta H^0_{\mathbf{k}}(t) - \Delta H^0_{\mathbf{k}}(t_0)$ along $H \rightarrow N \rightarrow \Gamma$ (represented by blue $\rightarrow$ white $\rightarrow$ red) as dots, with their areas proportional to $|\Delta H^0_{\mathbf{k}}(t)|$. At $t>t_s$, we see that that $H^0_{\mathbf{k}}(t)$ has reverted to its equilibrium condition ($t<t_0$).

Next, we examine the effects of el-el interactions by plotting the $\mathbf{k}$-resolved quasiparticle dispersion, defined as $H^\mathrm{tot}_{\mathbf{k}}(t) = \tilde H^0_{\mathbf{k}}(t) + H^\mathrm{corr}_{\mathbf{k}}(t)$, in Fig.~\ref{fig:nessi}b. We see that once $H^\mathrm{corr}_{\mathbf{k}}(t)$ is included, the green and magenta dots coincide with the equilibrium black line throughout the entire $\mathbf{k}$-path. Figure~\ref{fig:nessi}d confirms that the net change $\Delta H^\mathrm{tot}_{\mathbf{k}}(t) = H^\mathrm{tot}_{\mathbf{k}}(t) - H^\mathrm{tot}_{\mathbf{k}}(t_0)$ remains negligible at all times. In other words, the many-body correction $H^\mathrm{corr}_{\mathbf{k}}(t)$, which encapsulates the potential energy from el-el interactions, exactly cancels the kinetic energy shift $\Delta \tilde H^0_{\mathbf{k}}(t)$, preserving the total quasiparticle energy at each $\mathbf{k}$.

This discussion offers the clearest physical picture of el-el interactions and the physical implication of the many-body $GW$-correction. Electrons within the ionic potential of Na exhibit behavior characteristic of ideal harmonic oscillators: any increase in kinetic energy is exactly offset by a corresponding decrease in potential energy, thereby conserving the total energy. It also serves as a validation our numerical implementation: the total energy of our calculation is converged and conserved at all times. 

Finally, we note that with the inclusion of el-el interactions, the electron density at each $\mathbf{k}$-point, $n_\mathbf{k}(t)$, is no longer time-invariant (which is assumed in Eq.~\eqref{eq:Jind_main} for the noninteracting case). 
The nonlocal $\mathbf{q}$-dependent screened Coulomb interaction $W_{\mathbf q}(t,t')$ causes electrons to scatter between $\mathbf k$‐points, leading to the emergence of the paramagnetic current (Sec.~\ref{sec:paramagnetic_cur}). As a result, the single-particle occupations
\[
n_{\mathbf k}(t)=\langle c_{\mathbf k}^\dagger(t)\,c_{\mathbf k}(t)\rangle
\]
oscillate in time, as shown in Fig.~\ref{fig:nessi}c and in agreement with the continuity equation (Eq.~\eqref{eq:continuity_eq}). Note that this oscillation is larger than the numerical threshold within which  the total density is conserved in our calculations (Fig.~\ref{fig:nessi}f). 

Nonetheless, the Fourier components of the real‐space density
\[
n_{\mathbf q}(t)=\langle\hat n_{\mathbf q}(t) \rangle
=\sum_{\mathbf k}\langle c_{\mathbf k+\mathbf q}^\dagger(t)\,c_{\mathbf k}(t)\rangle
\]
remain zero for wave vectors $\mathbf q$ that are not reciprocal lattice vectors, $\mathbf G$,  i.e., $\mathbf{q} \notin \{\mathbf{G}\}$, due to the discrete translational symmetry enforced by the Hamiltonian. Meanwhile, the components at the Bragg peaks, $\langle \hat n_{\mathbf G}(t)\rangle$, remain locked to their equilibrium values. 
Here, $\big\langle \hat n_{\mathbf q/ \mathbf G}(t)\big\rangle$ denotes the expectation value of the density operator at time, $t$, in the quantum state at that instant. 
Physically, this means that despite the  optical excitation, there is no net modulation of the charge density in real space beyond its equilibrium value. The only spatial modulation in $n(\mathbf r,t)$ is the original crystal periodicity at equilibrium, with its amplitude locked to its equilibrium values at all times. Like in the non-interacting case, a curron has been excited without exciting any plasmon.

\subsection{Generalized d'Alembert wave equation} \label{sec:damped_dalembert_eq}

We postulate that el-el interactions can be incorporated phenomenologically by modifying the free-space d'Alembert wave equation (Eq.~\eqref{eq:dAlembert}) to include two damping terms:
\begin{align}
\epsilon_0 
 \frac{d^2 \mathbf{A}_{\mathrm{ind}}}{dt^2}  - \sigma \frac{d \mathbf{A}_{\mathrm{ind}}}{dt} 
&= \mathbf{J}_{\mathrm{ind}}(t) - L \frac{d \mathbf{J}_{\mathrm{ind}}}{dt}. \label{eq:modified_dAlembert}
\end{align}
The coefficient $\sigma$ is a conductivity-like term that resists changes in $\mathbf{A}_{\mathrm{ind}}$, analogous to how Ohmic conductivity describes resistive losses through $\mathbf{J} = \sigma \mathbf{E} = -\sigma \frac{d \mathbf{A}_{\mathrm{ind}}}{dt}$. The second coefficient, $L$, accounts for the delayed response of the induced current, analogous to the inductance term in $V = L\, dI/dt$ that resists changes in current. The combination $\sigma L$ has units of permittivity and can be interpreted as an interaction-renormalized $\epsilon_0$. In this way, Eq.~\eqref{eq:modified_dAlembert} becomes a dynamical extension of the free-space wave equation, describing a medium response that includes both memory and retardation effects. This formulation is particularly useful for extending one-body frameworks, such as time-dependent DFT, to account for interaction effects once $\sigma$ and $L$ are extracted from TT-$GW$.

For clarity, we focus on the regime after the external pulse has ended (i.e., Fig.~\ref{fig:int_ele}c, right) and assume time-translation invariance, such that the two-time dependence reduces to a single relative time. Substituting Eq.~\eqref{eq:Jind_retarded} into Eq.~\eqref{eq:modified_dAlembert} and Fourier transforming yields the frequency-domain dispersion relation:
\begin{equation}
\left(1 + iL\omega\right) {K}^{R}_{T}(\omega)
 = \epsilon_0\omega^2
 - i \sigma \omega.
\label{eq:dispersion_relation_in_freq}
\end{equation}
Solving for the current response, we obtain a constitutive relation for the interacting case:
\begin{equation}
{J}_{\mathrm{ind}}(\omega) = - {K}^{R}_{T}(\omega) {A}_{\mathrm{tot}}(\omega),
\nonumber
\end{equation}
with the kernel
\begin{equation}
{K}^{R}_{T}(\omega) = \dfrac{\epsilon_0 \omega^2 - i  \sigma \omega}{1 + i L \omega}.
\label{eq:kernel_in_freq}
\end{equation}
This response kernel contains four central contributions: 
(1) the instantaneous diamagnetic response ($D$); 
(2) the retarded paramagnetic response ($\chi_T^R$) from el-el interactions; 
(3) light-matter coupling via the $\epsilon_0\omega^2$ term from EM wave dynamics; and 
(4) interaction-induced damping via $\sigma$ and $L$.

To interpret the kernel $K^R_T(\omega)$, we examine four limiting cases: the noninteracting limit, the low-frequency and high-frequency regimes, and the resonance near $\omega = \omega_c$. In the noninteracting case ($\sigma = L = 0$), the kernel simplifies to $K(\omega) = \epsilon_0 \omega^2$, recovering the same instantaneous diamagnetic current, $\mathbf{J}_{\mathrm{ind}} = -\epsilon_0 \omega^2 \mathbf{A}_{\mathrm{tot}} = -D\, \mathbf{A}_{\mathrm{tot}}$, we obtained in Sec.~\ref{sec:nonintele} (Eq.~\eqref{eq:jind_main}). 

The phase $\phi_K = \arg K^R_T(\omega)$ controls the lag or lead between $\mathbf{J}_{\mathrm{ind}}$ and $-\mathbf{A}_{\mathrm{tot}}$. From Eq.~\eqref{eq:kernel_in_freq}, it evaluates to
\begin{align}
\phi_K (\omega)
&= \arctan\left[ \frac{ - (\sigma + \epsilon_0 L \omega^2)  }{ \omega ( \epsilon_0 - \sigma L) } \right] \nonumber\\
&= \arctan\left( \frac{ - \sigma }{ \epsilon_0 \omega} \right) - \arctan (\omega L),
\label{eq:phase_of_K}
\end{align}
demonstrating that $\sigma$ dominates at low frequencies, while $L$ dominates at high frequencies. Using the convention that time-dependent fields oscillate as $\sin(-\omega t + \mathrm{phase})$, a negative $\phi_K$ indicates that $\mathbf{J}_{\mathrm{ind}}$ leads $-\mathbf{A}_{\mathrm{tot}}$ in time, while a positive value indicates it lags.

In the low-frequency limit, $\omega \ll \omega_c$, expanding Eq.~\eqref{eq:kernel_in_freq} gives:
\[
{K}^{R}_{T}(\omega) \simeq -i \sigma \omega + (\epsilon_0 - \sigma L) \omega^2 + \mathcal{O}(\omega^3),
\]
where the leading term is imaginary and linear in $\omega$, consistent with resistive damping. Here, $\mathbf{J}_{\mathrm{ind}}$ leads by $90^\circ$.

In the high-frequency limit, $\omega \gg \omega_c$, we expand the kernel in $1/\omega$:
\[
{K}^{R}_{T}(\omega) \simeq -i \frac{\epsilon_0}{L} \omega + \frac{\epsilon_0 - \sigma L}{L^2} + \mathcal{O}\left(\frac{1}{\omega}\right),
\]
where the leading term is again purely imaginary and linear in frequency. This describes inductive behavior, but with the same phase structure as in the low-frequency limit: $\mathbf{J}_{\mathrm{ind}}$ again leads $-\mathbf{A}_{\mathrm{tot}}$ by $90^\circ$. Although the microscopic origin differs, with resistive damping at low frequency and inductive delay at high frequency, the phase relationship remains the same in both asymptotic regimes.

At resonance ($\omega = \omega_c$), the phase shift is alternatively derived in Appendix~\ref{sec_si:derive_wc}. When $\epsilon_0 > \sigma L$, $\phi_K < 0$ and $\mathbf{J}_{\mathrm{ind}}$ still leads $-\mathbf{A}_{\mathrm{tot}}$. The competing roles of $\sigma$ and $L$ can thus be viewed as a balance between resistive alignment with $\mathbf{E} \propto d\mathbf{A}_{\mathrm{tot}}/dt$ and inductive delay.

Fitting $\mathbf{A}_{\mathrm{ind}}$ and $\mathbf{J}_{\mathrm{ind}}$ gives $\sigma = 0.001~\mathrm{au}$ and $L = 3.6~\mathrm{au}$, which accurately reproduce the TT-$GW$ current in Fig.~\ref{fig:int_ele}c (right). These fitted parameters imply that inductive effects dominate in our metallic system. Solving Eq.~\eqref{eq:modified_dAlembert} with these values yields a resonant phase lead of $\phi_K = 32^\circ$ at $\omega_c$.

\subsection{Parallel RLC-circuit analogy}

To interpret physical implications of introducing the two damping terms into our damped d'Alembert wave equation, we organize the terms multiplying $A(\omega)$ in Eq.~\eqref{eq:dispersion_relation_in_freq} into a form resembling a parallel RLC-circuit:
\[
\left\{
   -\,\epsilon_0\,\omega^2
   + i\,\omega \left[ \sigma + L\,\boldsymbol{K}^{R}_{T}(\omega) \right]
   + \boldsymbol{K}^{R}_{T}(\omega)
\right\} A(\omega) = 0.
\]
This structure suggests that the dynamics can be modeled as a damped oscillator driven by the field, analogous to an RLC-circuit coupled to electromagnetic degrees of freedom (i.e., RLC+field model). The term $-\epsilon_0 \omega^2$ arises from the electromagnetic field and plays the role of a kinetic or inertial term. The imaginary term $i\omega \sigma$ corresponds to Ohmic dissipation from el-el interactions, acting as a resistive element. The memory-dependent term $i\omega L\,\boldsymbol{K}^{R}_{T}(\omega)$ introduces a frequency-dependent response analogous to an effective inductor. Finally, the static term $\boldsymbol{K}^{R}_{T}(\omega)$, appearing without derivatives, behaves as a restoring force, analogous to an inverse capacitance $1/C$.

The conservative part of the total energy is stored in three quadratic reservoirs:
\begin{equation} \label{eq:Utot}
U_{\text{tot}}(t) = U_C(t) + U_E(t) + U_L(t),
\end{equation}
where $U_C$ and $U_L$ represent capacitive and inductive energy storage in the electron fluid, while $U_E$ accounts for energy stored in the transverse electromagnetic field. In the absence of damping, energy oscillates coherently among these three components. The presence of $U_E$, which has no analog in a standard RLC-circuit, highlights the role of the field as a dynamical component that exchanges energy with the medium through its current response. We now examine each energy term in detail.

Firstly, the current response of the electron gas contributes a capacitive reservoir that exists even in the absence of interactions:
\[
U_C(t) = \tfrac{1}{2} D\, |A_{\mathrm{tot}}(t)|^2, \qquad D = {K}^{R}_{T}(\omega=0).
\]
The light-induced electric field, which opposes the vector potential that generated it, drives a charge displacement with a restoring force proportional to that displacement (see Appendix~\ref{sec_si:restoring_force}). The energy stored in this process is analogous to that in a capacitor, $U = \tfrac{1}{2} C V^2$, and can be obtained by integrating the instantaneous power transfer, $\mathbf{J}_{\mathrm{ind}} \cdot \mathbf{E}_{\mathrm{tot}}$, over time.

Secondly, the electromagnetic field itself stores energy in its transverse electric component:
\[
U_E(t) = \tfrac{1}{2} \epsilon_0 \left| \frac{d A_{\mathrm{tot}}}{dt} \right|^2,
\qquad
E_\mathrm{tot}(t) = -\frac{d A_{\mathrm{tot}}}{dt},
\]
which corresponds to the standard field energy density, $\tfrac{1}{2} \epsilon_0 |E(t)|^2$. This term accounts for electromagnetic inertia that resists rapid changes in the vector potential. Like $U_C$, it is present even in the noninteracting limit. In this limit, the system behaves as a C+field model, i.e., a C-type circuit coupled to and exchanging energy with a free electromagnetic field.

Thirdly, el-el interactions introduce an inductive energy reservoir:
\[
U_L(t) = \tfrac{1}{2} L\, \left| J_{\mathrm{ind}}(t) \right|^2, \qquad J_{\mathrm{ind}}(t) = -D\,A_{\mathrm{tot}}(t),
\]
which stores energy in the current itself, analogous to an inductor’s $\tfrac{1}{2} L I^2$. This term reflects how interactions resist changes in current, leading to delayed backflow. It is absent in the noninteracting limit, where $\chi^R_T(t,t') = 0$, and encodes the memory inherent in the many-body response.

Electron-electron interactions also introduce damping, represented by the combined rate of $\sigma + L {K}^{R}_{T}(\omega=0)$ (see Sec.~\ref{sec:damped_dalembert_eq}). Differentiating Eq.~\eqref{eq:Utot} and substituting the equations of motion yields the instantaneous power loss:
\[
P_{\text{loss}}(t) = - \frac{dU_{\text{R}}}{dt}
= \left[\sigma + L {K}^{R}_{T}(0) \right]\,\left|\frac{d A_{\mathrm{tot}}}{dt}\right|^{2},
\]
where $U_R(t)$ is the cumulative energy irreversibly dissipated through Ohmic scattering ($\sigma$) and interaction-induced damping ($L K_T^R(0)$). This is the analog of the resistive term $R I^2$ in an RLC-circuit.

In this analysis, we approximate the frequency-dependent kernel $\chi_T^R(\omega)$ by its zero-frequency limit, $K_T^R(0)$, i.e., a static approximation. While this captures the correct energy loss rate in steady state, it neglects the retarded part of the kernel and is valid only when $\chi_T^R(\omega)$ varies slowly over the frequency range of interest. Nonetheless, it offers conceptual clarity and an intuitive picture to interpret our generalized d'alembert equation. Upon the inclusion of With el-el included, the system is generalized from a C+field to an RLC+field model, where the resistive ($\sigma$) and inductive ($L$) terms emerge from many-body physics.

\section{Conclusion} \label{sec:conclusion}

In this work, we have analyzed the generation and properties of currons in systems of both noninteracting and interacting electrons under non-equilibrium conditions. Our study reveals fundamental distinctions between a plasmon and a curron, while highlighting the impact of el-el interactions. 

We found that the ratio of the induced current density relative to the total vector potential is the negative Drude weight, which is a material property that depends only on the electronic structure of the material. The curron frequency, like the plasmon frequency, is the natural frequency of the electron density. What differentiates a curron from a plasmon is the nature of the external driver. While the former is driven by transverse electromagnetic oscillations of light, the latter is driven by longitudinal Coulomb interactions between the electrons. 

In order to account for el-el interactions and self-energy corrections, we employed the TT-$GW$ formalism. Our results show that while the longitudinal electric field due to el-el interactions is orders of magnitude larger than the transverse electric field of light and that the former can lead to significant renormalization of the curron amplitude that the latter cannot, they are unable to excite and sustain a curron. However, dynamical screening effects due to el-el interactions lead to a renormalization of the magnitude of the induced current density and a phase lag of the vector potential relative to the current that induced it. Phenomenologically, the effects of el-el interaction can be modeled by introducing two damping terms in the  d'Alembert wave equation, using an inductance-like term and a resistive-like term, such that energy exchange between the energy stored in the material through el-el interactions and energy from the electric field component of the light can be modeled like an RLC+field model. By storing the potential energy, el-el interactions resist changes to the current density.

Our analysis of time-dependent external fields suggests that the interplay between interactions and driving fields can lead to nontrivial modifications in current generation. We anticipate that this new perspective will illuminate the interplay between \emph{longitudinal} and \emph{transverse} fields in many-body systems and stimulate further theoretical and experimental investigations into nontrivial electromagnetic response in novel materials, with implications for the design of quantum materials and devices operating under strong external drives, such as ultrafast electronics and light-induced phase transitions. Future work may explore extensions to more complex correlated systems and incorporating vertex corrections beyond $GW$. 

Finally, experimental verification of the predicted renormalization effects in driven electronic systems would provide further validation of the theoretical framework presented here. While early optical experiments~\cite{Sutherland1967, Sutherland1969} on sodium films reported features near the plasmon frequency, these measurements lacked the resolution, polarization control, and momentum sensitivity needed to identify the nature of the excitation. Today, experimental detection of currons can benefit significantly from modern synchrotron-based spectroscopic methods. Finite-momentum electron energy-loss spectroscopy and resonant inelastic x-ray scattering can directly probe the distinctive dispersion and energy scale of transverse collective modes relative to conventional longitudinal plasmons. Polarization- and angle-resolved vacuum-ultraviolet (VUV) reflectometry, exploiting polarized incident radiation, offers a complementary route to isolate the transverse dielectric function and identify the characteristic spectral signature predicted near the curron resonance. Additionally, time-resolved VUV spectroscopy using attosecond high-harmonic pulses could capture the real-time dynamics of current-driven excitations, thus directly testing the predictions of the two-time $GW$ framework. These approaches collectively represent powerful avenues for experimentally observing and characterizing currons in real quantum materials.

\begin{acknowledgments}
C.-S.O and O.E. acknowledge support by the Swedish Research Council (Vetenskapsrådet, VR), the European Research Council (ERC) (synergy grant FASTCORR, project no. 854843). O.E. acknowledges support from the Wallenberg Initiative Materials Science for Sustainability (WISE) funded by the Knut and Alice Wallenberg Foundation (KAW), support from VR, eSSCENCE and STandUPP. H.U.R.S. acknowledges financial support from the VR grant number 2024-04652. The computations were enabled by resources provided by the National Academic Infrastructure for Supercomputing in Sweden (NAISS/SNIC) partially funded by VR. 
\end{acknowledgments}

\appendix
\section{DFT Calculations} \label{sec_si:dft}

The density-functional theory (DFT) calculations were performed using the \texttt{Quantum Espresso}~\cite{Giannozzi2009} package. The local-density approximation (LDA) was used for the electron exchange and correlation energy. A scalar-relativistic ONCVPSP pseudopotential~\cite{Hamann2013} for Na obtained from the PSEUDODOJO project~\cite{vanSetten2018} is used for the calculation. The plane-wave cutoff for the DFT calculation was set to 100~Ry and 40~Ry for the plane-wave expansion of the wavefunctions. Integration over the Brillouin zone was calculated on a $\mathbf{k}$-grid of $24\times24\times24$.

The DFT band structure calculations were performed for the primitive unit cell of the experimental~\cite{Wyckoff1963} structure of body-centered cubic Na, where $ a = b = 4.2906 $~\AA. The special k-points along the high-symmetry path are listed in Table~\ref{tab:kpoints}.
Coordinates are expressed in units of the primitive bcc reciprocal vectors that correspond to the direct lattice:
\[
(-\tfrac12,\,\tfrac12,\,\tfrac12)\,a,\quad(\tfrac12,\,-\tfrac12,\,\tfrac12)\,a,\quad(\tfrac12,\,\tfrac12,\,-\tfrac12)\,a.
\]

\begin{table}[h!]
    \centering
    \caption{Special $\mathbf{k}$-points used in the band structure calculations. 
    The coordinates $k_1$, $k_2$, and $k_3$ are fractional components of the primitive reciprocal lattice vectors $\mathbf{b}_1$, $\mathbf{b}_2$, and $\mathbf{b}_3$, respectively. The corresponding Cartesian components in the conventional cubic basis are denoted by $k_x$, $k_y$, and $k_z$.}    
    \label{tab:kpoints}
    \begin{tabular}{|c|ccc|ccc|}
        \hline
        \multirow{2}{*}{Label} & \multicolumn{3}{c|}{Primitive} & \multicolumn{3}{c|}{Conventional} \\
                               & $k_1$ & $k_2$ & $k_3$ & $k_x$ & $k_y$ & $k_z$ \\
        \hline
        $\Gamma$ & 0    &  0   &  0   &  0   & 0   & 0   \\
        H        & 1/2  & -1/2 &  1/2 &  0   & 1   & 0   \\
        N        & 0    &  0   &  1/2 &  1/2 & 1/2 & 0   \\
        $\Gamma$ & 0    &  0   &  0   &  0   & 0   & 0   \\
        P        & 1/4  &  1/4 & 1/4  &  1/2 & 1/2 & 1/2 \\
        H        & 1/2  & -1/2 & 1/2  &  0   & 1   & 0   \\
        \hline
    \end{tabular}
\end{table}

\section{Symmetry constraints on $v_x$ along the high-symmetry path} \label{sec_si:vx=0}

This Appendix expands the discussion of symmetry-enforced zeros of the band velocity by examining the $x$-component along the path specified in Table~\ref{tab:kpoints}:
\begin{align*}
\Gamma \rightarrow 
H \rightarrow
N \rightarrow 
\Gamma \rightarrow 
P \rightarrow 
H.
\end{align*}

The group velocity in a single-band model is given by $v_x=\hbar^{-1}\,\partial\varepsilon_{\mathbf{k}}/\partial k_x$ (Eq.~\eqref{eq:velocity_k_main}). Its expectation value vanishes only when at least one operation in the little co-group maps $k_x \mapsto -k_x$ while leaving $\mathbf{k}$ invariant modulo a reciprocal lattice vector. Such an operation sends $v_x \mapsto -v_x$ without altering the Bloch state. Since the phase factors introduced by the symmetry cancel in the expectation value, this leads to the identity $\langle v_x \rangle = -\langle v_x \rangle$, which implies that $\langle v_x \rangle = 0$. In the absence of any such sign-reversing symmetry, $v_x$ is unconstrained by symmetry and is determined purely by the dispersion.

Na belongs to the paramagnetic cubic space group \textit{Im$\bar{3}$m} (No.~229). At $\Gamma$ and $H$, the full cubic point group $O_h$ ($m\bar{3}m$) contains a mirror plane that maps $x \mapsto -x$, so $v_x$ vanishes at both points. The entire segment $\Gamma \rightarrow H$ lies on the $\Delta$ line, which has little co-group $C_{4v}$ ($4m.m$). One of its vertical mirrors lies in the $yz$-plane, reverses $x$, and leaves each point on the line invariant. Therefore, $v_x$ must vanish throughout this branch.

The point $N$ has orthorhombic site symmetry $D_{2h}$ ($m.mm$), which includes two perpendicular mirror planes. One of them is the plane $x=0$, which flips $x \mapsto -x$ and leaves $N$ fixed, thereby enforcing $v_x = 0$ at that point. The two adjoining segments, $H \rightarrow N$ and $N \rightarrow \Gamma$, lie on the $G$- and $\Sigma$-lines. Each line has little co-group $C_{2v}$ ($m.m2$). Every symmetry operation in the little co-group must leave each point on the line invariant modulo a reciprocal lattice vector, so the two vertical mirrors and the two-fold rotation axis all contain the $k$-direction of the line (which, in \textit{Im$\bar{3}$m}, lies along a $\langle110\rangle$ direction). Neither mirror reverses $x$ without displacing the point off the line, and the two-fold rotation exchanges $x$ and $y$ rather than flipping either. Consequently, $v_x$ is unconstrained along these segments and is determined by the band structure.

After returning to $\Gamma$, the path proceeds to $P$ along the $J$-line, which has little co-group $C_s$ ($..m$). Only a single mirror survives on this segment. Since the $J$-line lies along the $[111]$ direction, any operation in the little co-group must preserve that direction, i.e., it must leave all vectors satisfying $x = y = z$ invariant. The surviving mirror therefore cannot reverse $x$ independently, because such an operation would also need to reverse $y$ and $z$ to preserve the $[111]$ direction, which no mirror does. A mirror that flips only $x$ would not preserve $\mathbf{k}$ along the line and would thus map it outside the star of that line, excluding the operation from the little co-group. As a result, symmetry does not constrain $v_x$ along the $J$-line.

At the zone corner $P$, the little co-group is the tetrahedral group $T_d$ ($\bar{4}3m$), which lacks any operation that reverses $x$, so $v_x$ can remain finite. The final stretch $P \rightarrow H$ also lies on a $G$-type line with little co-group $C_{2v}$; the earlier argument applies again: $v_x$ is unrestricted until the path reaches $H$, where symmetry enforces $v_x = 0$.

In summary, $v_x$ must vanish at the special points $\Gamma$, $H$, and $N$, and everywhere along the $\Delta$ line from $\Gamma$ to $H$. Along all other segments of the path, $H \rightarrow N$, $N \rightarrow \Gamma$, $\Gamma \rightarrow P$, and $P \rightarrow H$, the symmetry of \textit{Im$\bar{3}$m} does not constrain $v_x$, which remains free to vary according to the band dispersion.

\section{Construction of Downfolded Wannier Hamiltonians} \label{sec_si:w90}

The Wannier Hamiltonians were constructed using DFT wavefunctions calculated non-self-consistently on a $24\times24\times24$ $\mathbf{k}$-grid, with the 3$s$ pseudoatomic wavefunction as the initial guess for the Wannier basis states.

To account for the screening effects of the unoccupied states that were removed by downfolding, we employed the constrained random phase approximation (cRPA) within the static limit ($\omega = 0$)~\cite{Aryasetiawan2004, Nakamura2021}. The cRPA calculations were performed using \texttt{REPSPACK}~\cite{Nakamura2021} in combination with \texttt{wan2respack}~\cite{Kurita2023}, which interfaces with \texttt{Quantum ESPRESSO}. We included screening using 100 unoccupied bands on $24\times24\times24$ $\mathbf{k}$-grid.

\begin{figure}[htbp]
    \centering
    \includegraphics[width=\linewidth]{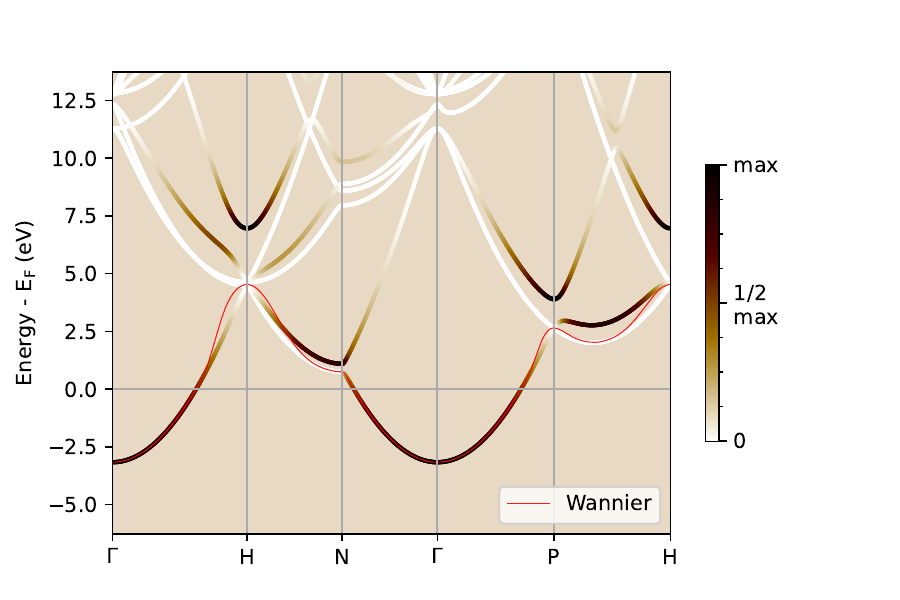}
    \caption{Wannier tight-binding band structure (in red) superimposed on the DFT band structure. 
    The DFT band structure is plotted with a color map corresponding to its projection onto the 3$s$ pseudoatomic wavefunction.}
    \label{fig:w90_bs}
\end{figure}

The screened interaction was modeled using the Fourier-transformed Yukawa potential,
\begin{equation}
    V^{\mathrm{cRPA}}(q) = \frac{1}{V_0 \epsilon_0} \frac{k e^2}{q^2 + \lambda^2},
\nonumber
\end{equation}
where $ \epsilon_0 $ is the permittivity of free space, $ e $ is the elementary charge, $ q $ is the wavevector, $ \lambda $ is the inverse screening length, and $ V_0 $ is a dimensionless scaling factor. 
By fitting the static screened interaction obtained from cRPA to a Yukawa potential, we determine $ \lambda = 0.075~\si{bohr^{-1}} $ and $ V_0 = 10.8 $, which are used in our TT-$GW$ calculations.

\section{Derivation of the Restoring Force Density, $\mathbf{f}_\mathrm{res}$, of a  Homogeneous $\mathbf{A}_{\mathrm{ext}}$ }  \label{sec_si:restoring_force}

Starting from Newton's second law of motion, the force on a single electron is given by.
\begin{equation} \label{eq_si:newton_2nd_law}
m_{\mathrm{eff}} \frac{d \mathbf{v}}{dt} = q \mathbf{E}_{\mathrm{tot}}(t).
\end{equation}
Multiplying both sides with $n_e$ to obtain the force density, $\mathbf{f}(t)$, and substituting in the Faraday's law, $\mathbf{E}_{\mathrm{tot}}(t) = -\frac{d \mathbf{A}_{\mathrm{tot}}}{dt}$ (Eq.~\eqref{eq:Faraday's_law}), and $\mathbf{J}_{\mathrm{ind}}(t)= n_e q \mathbf{v}(t)$, we get
\begin{equation}
\frac{m_{\mathrm{eff}}}{q} \frac{d \mathbf{J}_{\mathrm{ind}}}{dt} = -n_e q \frac{d \mathbf{A}_{\mathrm{tot}}}{dt}
\end{equation}
Differentiating both sides, 
\begin{equation}
\frac{m_{\mathrm{eff}}}{q} \frac{d^2 \mathbf{J}_{\mathrm{ind}}}{dt^2} = -n_e q \frac{d^2 \mathbf{A}_{\mathrm{tot}}}{dt^2}
\end{equation}
and substituting in the d'Alembert wave equation (Eq.~\eqref{eq:dAlembert}), we get
\begin{equation}
\frac{m_{\mathrm{eff}}}{q} \frac{d^2 \mathbf{J}_{\mathrm{ind}}}{dt^2} = -\frac{n_e q}{\epsilon_0}\,\mathbf{J}_{\mathrm{ind}}(t),
\label{eq_si:intermediate_restoringforce}
\end{equation}
or equivalently,
\begin{equation}
\frac{d^{2}J_{\mathrm{ind}}}{dt^{2}} + \omega_{c}^{2}\,J_{\mathrm{ind}}(t) = 0,
\qquad
\omega_{c}^{2} \equiv \frac{n_{e}q^{2}}{m_{\!\text{eff}}\varepsilon_{0}} = \frac{D}{\epsilon_{0}}.
\label{eq:harmonic}
\end{equation}
This is the equation of a simple harmonic oscillator with angular frequency $\omega_{c}$ (that is the curron frequency as defined in Eq.~\eqref{eq:wc_main}). The general solution for $J_{\mathrm{ind}}(t)$ is
\begin{equation}
J_{\mathrm{ind}}(t) = J_{0}\cos\left[\omega_{c}(t - t_{0}) + \phi\right], \label{eq_si:Jind_analytic}
\end{equation}
where $J_{0}$ and $\phi$ are determined by initial conditions.

The restoring force density is given by the integral of the right-hand side of Eq.~\eqref{eq_si:intermediate_restoringforce}. Further substituting in the general solution of $\mathbf{J}_{\mathrm{ind}}(t)$ from Eq.~\eqref{eq_si:Jind_analytic}, we get,
\begin{align}
\mathbf{f}_{\mathrm{res}}(t)
&= -\frac{n_e q}{\epsilon_0} \int_{t_0}^t \mathbf{J}_{\mathrm{ind}}(t')\, dt' \nonumber \\
&= -\frac{n_e q}{\varepsilon_{0}}\,
    \frac{J_0}{\omega_c}\,
    \sin\!\bigl[\omega_c(t-t_0)+\phi\bigr]\,
    \hat{\mathbf{x}}
\nonumber \\[4pt]
&= -\,n_e m_{\mathrm{eff}}\omega_c^{2}\,\boldsymbol{\xi}(t),
\label{eq_si:force_density}
\end{align}
where $\boldsymbol{\xi}(t)$ is the charge displacement given by, 
\begin{align}
\boldsymbol{\xi}(t) 
&= \frac{1}{n_e q} \int_{t_0}^t \mathbf{J}(t)  dt \nonumber  \\
&= \frac{J_0}{n_e q \omega_{c}} \sin\left[\omega_{c}(t - t_{0}) + \phi\right] \mathbf{\hat{x}}.
\end{align}
Consequently, $\mathbf{f}_{\mathrm{res}}(t)$ is non-zero when $\mathbf{J}_{\mathrm{ind}}(t) = 0$ (at maximum displacement), consistent with harmonic oscillation dynamics where force and current are $90^\circ$ out of phase.

Summarizing, the induced electric field arises in response to changes in the vector potential, opposing the change in vector potential that created it and driving charge motion. The resulting charge displacement creates an electrostatic restoring force proportional to the charge displacement, $\boldsymbol{\xi}(t)$. This force density, $\mathbf{f}_{\mathrm{res}}(t)$, (Eq.~\eqref{eq_si:force_density}) opposes the displacement and restores equilibrium.

\section{Derivation of $\omega_c$ for interacting electrons} \label{sec_si:derive_wc}

We start  from the modified d'Alembert wave equation (Eq.~\eqref{eq:modified_dAlembert} of Main Text), in which the effects of el-el interactions are represented by the $L$ and $\sigma$ damping terms:
\begin{align} \label{eq:modified_dAlembert_si}
\epsilon_0 
 \frac{d^2 \mathbf{A}_{\mathrm{ind}}}{dt^2}  -\sigma \frac{d \mathbf{A}_{\mathrm{ind}}}{dt} 
&= \mathbf{J}_{\mathrm{ind}}(t) - L \frac{d \mathbf{J}_{\mathrm{ind}}}{dt}
\end{align}

Next, we assume that the induced vector potential $A_{\mathrm{ind}}$ and the induced current density $J_{\mathrm{ind}}$ vary harmonically with time (as in Sec.~\ref{sec:linearresponselimit}),
\begin{align}
    A_{\mathrm{ind}}(t) &= A_0 e^{-i\omega t}, \\
    J_{\mathrm{ind}}(t) &= J_0 e^{-i\omega t + i\tilde \phi},
\end{align}
where $\omega$ is the angular frequency,  $\tilde \phi$ is the phase of $J_{\mathrm{ind}}(t)$ relative to $A_{\mathrm{ind}}(t)$. Since we know from our study with noninteracting electrons that $J$ opposes $A$ due to Lenz's law, we will bake the negative sign of Eq.~\eqref{eq:jind_main} (of the Main Text) into $\tilde \phi$ by letting $\tilde \phi = \phi_K + \pi$, such that $\phi_K$ will then be phase shift of the kernel, i.e., phase of $J_{\mathrm{ind}}(t)$ relative to $-A_{\mathrm{ind}}(t)$. 

Now, the first and second derivatives of $A_{\mathrm{ind}}$ are given by,
\begin{align}
\frac{d A_{\mathrm{ind}}}{dt} &= -i\omega A_0 e^{-i\omega t}, \label{eq:dAdt_si}\\
\frac{d^2 A_{\mathrm{ind}}}{dt^2} &= -\omega^2 A_0 e^{-i\omega t}. \label{eq:d2Adt2_si}
\end{align}
and the first derivative of $J_{\mathrm{ind}}$ is given by,
\begin{align}
\frac{d J_{\mathrm{ind}}}{dt} = -i\omega J_0 e^{-i(\omega t + \tilde \phi)}. \label{eq:dJdt_si}
\end{align}
Substituting Eqs.~\eqref{eq:dAdt_si}, \eqref{eq:d2Adt2_si} and \eqref{eq:dJdt_si} into Eq.~\eqref{eq:modified_dAlembert_si}, we get,
\begin{align}
 (-\epsilon_0 \omega^2 + i \sigma \omega) A_0 e^{-i \omega t} &= ( 1 + i L \omega ) J_0  e^{-i (\omega t + \tilde \phi)} \nonumber\\
- \epsilon_0 \omega^2 + i \sigma \omega  &= \frac{J_0} {A_0} e^{i \tilde \phi} ( 1 + i L \omega ). \label{eq:LC_harmonic}
\end{align}
Defining $\omega_0$ as the bare curron frequency for noninteracting electrons (see Eq.~\eqref{eq:curron_freq_nonint} of Main Text),
\begin{equation}
\omega_0^2 = \frac{J_0}{\epsilon_0 A_0},
\end{equation}
we rewrite Eq.~\eqref{eq:LC_harmonic} as:
\begin{align}
- \epsilon_0 \omega^2 + i \sigma\omega &= \epsilon_0 (1 + i L\omega) \omega_0^2 e^{-i\tilde \phi} \nonumber\\
-\epsilon_0 \omega^2 + i \sigma\omega &= 
\begin{aligned}[t]
&\epsilon_0 \omega_0^2 (\cos\tilde \phi - L\omega \sin\tilde \phi) \\
&\quad+ i \epsilon_0\omega_0^2 (\sin\tilde \phi + L\omega \cos\tilde \phi).
\end{aligned}
\end{align}

At resonance, the imaginary part vanishes. Therefore, we can solve for $\omega=\omega_c$ and $\tilde \phi=\phi_c$ by equating the real and imaginary parts to obtain a system of equations, 
\begin{align}
\text{Real:} \quad & -\omega_c^2 = \omega_0^2 (\cos\tilde \phi - L\omega_c \sin\tilde \phi), \label{eq:real_part_wc}\\
\text{Imaginary:} \quad & \sigma\omega_c = \epsilon_0\omega_0^2 (\sin\tilde \phi + L\omega_c \cos\tilde \phi). \label{eq:imag_part_wc}
\end{align}
First, we solve for $\omega_c$ by eliminating $\tilde \phi$. Squaring Eqs.~\eqref{eq:real_part_wc} and \eqref{eq:imag_part_wc} and adding them together, we get,
\begin{align}
\omega_c^4 + \frac{\sigma^2}{\epsilon_0^2}\omega_c^2 &=  \omega_0^4 (1+L^2\omega_c^2)\,
\end{align}
which is a quadratic equation in $\omega_c^2$, with its positive solution (since $\omega^2 \ge 0$) being,
\begin{equation}
\omega_c^2 = \frac{-\frac{\sigma^2}{\epsilon_0^2}+L^2\omega_0^4 + \sqrt{\left(\frac{\sigma^2}{\epsilon_0^2}-L^2\omega_0^4\right)^2+4\omega_0^4}}{2}.
\end{equation}
Next, we solve for $\tilde \phi$ by dividing Eq.~\eqref{eq:imag_part_wc} by Eq.~\eqref{eq:real_part_wc}, 
\begin{equation} \label{eq:tan_phi}
\tan\tilde  \phi = \frac{ - (\sigma + \epsilon_0 L \omega_c^2)  }{ \omega_c ( \epsilon_0 - \sigma L) } ,
\end{equation}
where the following particular solution for $\tilde \phi$ solves Eq.~\eqref{eq:modified_dAlembert_si}, gives
\begin{align}
\phi_K (\omega_c) = \tilde{\phi} - \pi 
&= 
\arctan\left[ \frac{ - (\sigma + \epsilon_0 L \omega_c^2)  }{ \omega_c ( \epsilon_0 - \sigma L) } \right] \nonumber\\
&= \arctan\left( \frac{ - \sigma }{ \epsilon_0 \omega_c} \right)
- \arctan (\omega_c L).
\label{eq:phi_sol}
\end{align}

Summarizing, we see that el-el interaction has two main renormalization effects on the curron frequency, $\omega_c$. First, el-el interactions scale the bare $\omega_0$ frequency by a factor of $-(\cos\tilde \phi + L\omega \sin\tilde \phi)$ (Eq.~\eqref{eq:real_part_wc}).  Second, the condition $L \gg \sigma$ indicating that retardation effects (quantified by $L$) dominate over resistive damping ($\sigma$), results in a positive phase shift $\phi$ where $\mathbf{J}_{\mathrm{ind}}$ leads $-\mathbf{A}_{\mathrm{tot}}$ by $\phi_K$ (Eq.~\eqref{eq:phi_sol}) due to delayed feedback from interaction-induced fields. Phenomenologically, this represents an inductive-like effect where the inherent opposition to current changes delays $\mathbf{A}_{\mathrm{ind}}$ buildup through storage of electric field energy (distinct from magnetic energy storage in conventional inductors), coupled with minor damping from $R$.

\section{Time-dependent TT-$GW$ calculations} \label{sec_si:tdgw}

Our TT-$GW$ for Na is solved for a unit cell on a $24\times24\times24$ $\mathbf{k}$-grid. The time step of $h = 0.004$~\si{as} was used and 4000 points on the Matsubuara branch were used to discretize imaginary time. Inverse temperature, $\beta=\frac{1} {k_\mathrm{B} T}$ is set to 50.0~\si{Ha^{-1}}. For each self-consistent step, $G(t,t')$ was mixed with 0.5 of the previous iteration, while the $W(t,t')$ was mixed with 0.6 of the previous iteration, and were converged to the threshold of $10^{-10}$~\si{au}. The total energies of all calculations are conserved up to an energy drift of less than $10^{-5}$~\si{eV} per unit cell. Our results are also converged with respect to the charge density to a threshold of less than $10^{-6}$ electron per unit cell. The calculations were performed up to 170~\si{as}. The plots for $E_{\mathrm{ind}}(t)$, $A_{\mathrm{ind}}(t)$ and $J_{\mathrm{ind}}(t)$ in Fig.~\ref{fig:int_ele} were extrapolated by fitting their steady-state behavior to sinusoidal curves.

With the use of an effective Hamiltonian ($H^0_{\mathbf{k}} (t)$) that is downfolded from the Kohn-Sham Hamiltonian ($H^{\mathrm{DFT}}_{\mathbf{k}}$), three comments are in order. Firstly, since unoccupied high-energy bands are removed in the effective model, we account for screening due to these states using the constrained RPA approximation~\cite{Aryasetiawan2004, Nakamura2021, Kurita2023} (see Appendix \ref{sec_si:w90}).

Secondly, maximally localized Wannier functions constructed from a finite set of Bloch bands form a complete basis for the corresponding low-energy subspace of the Hilbert space, but not for the full Hilbert space. Truncating the conduction manifold removes high-energy states that enter the sums for 
the perturbed Green function $G^{(1)}$; the missing weight carries over to the screened interaction $W$ and the self-energy $\Sigma$. One should therefore monitor the convergence of $W$ and $\Sigma$ with respect to the size of the Wannier window. This limitation can be alleviated with Wannier-Function Perturbation Theory~\cite{Lihm2021} or by introducing a hybridization self-energy $\Delta(\mathbf{k};t,t')$ that embeds the omitted bands as a bath, in close analogy to the treatment of hybridization in dynamical mean-field theory (DMFT).
Since virtual transitions through the omitted states are not included, the self-energy reported here is expected to be overestimated.

Thirdly, since $H^{\mathrm{DFT}}_{\mathbf{k}} = - \frac{\hbar}{2m} \boldsymbol{\nabla}^2 + V^{\mathrm{ion}}_{\mathbf{k}} + V^{\mathrm{H}}_{\mathbf{k}} + V^{\mathrm{xc}}_{\mathbf{k}}$, where $V^{\mathrm{ion}}_{\mathbf{k}}$ is ionic potential, $V^{\mathrm{H}}_{\mathbf{k}}$ is Hartree potential and $V^{\mathrm{xc}}_{\mathbf{k}}$ is exchange-correlation potential, $H^{\mathrm{DFT}}_{\mathbf{k}}$ already contain effects of the el-el interaction through  $V^{\mathrm{H}}_{\mathbf{k}}$ (mean-field effects) and $V^{\mathrm{xc}}_{\mathbf{k}}$ (exchange-correlation effects). These terms have to be removed to avoid double-counting and in this work, they are replaced by their time-dependent counterpart calculated within the Green's function framework, namely $V^{\mathrm{H}}_{\mathbf{k}}(t)$ and $H^\mathrm{corr}_{\mathbf{k}}(t)$. If the basis set is complete, the DFT $V^{\mathrm{H}}_{\mathbf{k}}$ will be equal to its Green's function equivalent, $V^{\mathrm{H}}_{\mathbf{k}}(t=t_0)$. Since we do not have a complete basis set, we correct for this by renormalizing the bandwidth of $H^0_{\mathbf{k}}(t)$ and $H^\mathrm{tot}_{\mathbf{k}}(t)$ to match that of $H^{\mathrm{DFT}}_{\mathbf{k}}$ (in Fig.~\ref{fig:nessi}), taking advantage of the fact that the DFT $V^{\mathrm{xc}}_{\mathbf{k}}$ already captures the many-body interaction reasonably well at equilibrium~\cite{Hedin1970, Hybertsen1986, Jensen1985, Northrup1987, Lyo1988, Northrup1989}.

\bibliography{main} 
\end{document}